\documentclass[prb,superscriptaddress,showpacs,floatfix,tightenlines,10pt,twocolumn]{revtex4}
\usepackage{amssymb}
\usepackage{amsmath}
\usepackage{graphicx}

\begin{document}

\title{Transport gap and hysteretic behavior of the Ising quantum Hall
ferromagnets in Landau levels $\left\vert N\right\vert >0$ of bilayer
graphene}
\author{Wenchen Luo}
\affiliation{D\'{e}partement de physique, Universit\'{e} de Sherbrooke, Sherbrooke, Qu%
\'{e}bec, J1K 2R1, Canada}
\author{R. C\^{o}t\'{e} }
\affiliation{D\'{e}partement de physique, Universit\'{e} de Sherbrooke, Sherbrooke, Qu%
\'{e}bec, J1K 2R1, Canada}

\begin{abstract}
The chiral two-dimensional electron gas in Landau levels $\left\vert
N\right\vert >0$ of a Bernal-stacked graphene bilayer has a
valley-pseudospin Ising quantum Hall ferromagnetic behavior at odd filling
factors $\nu _{N}=1,3$ of these fourfold degenerate states. At zero
interlayer electrical bias, the ground state at these fillings is spin
polarized and electrons occupy one valley or the other while a finite
electrical bias produces a series of valley pseudospin-flip transitions. In
this work, we extend the Ising behavior to chirally-stacked multilayer
graphene and discuss the hysteretic behavior of the Ising quantum Hall
ferromagnets. We compute the transport gap due to different excitations:
bulk electron-hole pairs, electron-hole pairs confined to the coherent
region of a valley-pseudospin domain wall, and spin or valley-pseudospin
skyrmion-antiskyrmion pairs. We determine which of these excitations has the
lowest energy at a given value of the Zeeman coupling, bias, and magnetic
field.
\end{abstract}

\date{\today }
\maketitle

\section{INTRODUCTION}

The chiral two-dimensional electron (C2DEG)\ gas in bilayer graphene (BLG)
in a quantizing magnetic field exhibits a rich variety of broken-symmetry
states in mean-field theory. These states have been extensively studied\cite%
{barlasreview} in the zeroth-energy Landau level $N=0$ which is eightfold
degenerate when valley $\left( K_{\pm }\right) $ and spin $\left( \sigma
=\pm 1\right) $ degrees of freedom are counted because of the presence of an
extra orbital degree of freedom $n=0,1$\cite{graphenereview}$.$ These broken
symmetry states are best described using the pseudospin language where a
valley pseudospin up(down)\ is associated with valley $K_{+}\left(
K_{-}\right) $ and an orbital pseudospin up(down)\ is associated with the
orbital $n=0(n=1).$ In this language, the broken-symmetry states are spin,
valley, or orbital quantum Hall ferromagnets (QHFs). They are gapped, have
charged quasiparticle excitations, and support an integer quantum Hall effect%
\cite{barlasreview}.

The effect of the Coulomb interaction on the higher Landau levels $%
\left\vert N\right\vert >0$ has so far received less attention. In a recent
paper\cite{luo}, we have shown that the C2DEG in these levels behaves as a
valley Ising QHF\ when the filling factor of these levels (which are
fourfold degenerate) is odd, i.e. when $\nu _{N}=1,3.$ The C2DEG at these
fillings is spin polarized and the two valley states with valley pseudospin $%
P_{z}=\pm \frac{1}{2}$ are degenerate at zero and at some finite values of
the bias $\Delta _{B}$ (the electrical potential difference between the two
layers). For $\nu _{N}=2,$ transitions occur between states with total spin
and total valley pseudospin $\left( S_{z},P_{z}\right) =\left( 0,\pm
1\right) ,\left( 1,0\right) .$ Changing the bias can induce first-order
transitions between the different spin and valley-pseudospin ground states.
The phase diagram for different Landau levels shows a marked difference
between the positive $N\geq 1$ and negative $N\leq -1$ Landau levels. The
Ising behavior of the higher Landau levels contrasts with that of Landau
level $N=0$ where valley, spin, or orbital QHF can all have a $U\left(
1\right) $ symmetry\cite{jules}.

Ising QHFs have been studied previously both theoretically\cite{quinn,
jungwirth,jungwirth2,jungwirth3} and experimentally\cite{daneshvar,piazza}
in a variety of semiconductor quantum wells or bilayer systems. A complete
classification of the QHF states, including isotropic, easy-axis, and
easy-plane QHFs has been derived for a model of two nearby, infinitely
narrow, two-dimensional layers\cite{jungwirthprb}. A well-known case of
Ising QHF occurs when two levels with different electrical subbands or
Landau level indices are brought to degeneracy by tuning the electrical bias
or the Zeeman coupling at even filling factor. The \textit{valley}%
-pseudospin Ising transitions in the higher Landau levels, however, occur at
odd filling factors and are of a slightly different nature: they do not
occur near Landau level crossings. As in other Ising QHFs, however, the
exchange part of the Coulomb interaction energy plays a major role as it
competes with the noninteracting energy. What is special with BLG is that an
electronic state is described by a four-component spinor and so consists of
a superposition of four states with different orbital wave functions. As the
bias is varied, the coefficients of this superposition are modified and so
is the Coulomb exchange interaction which is orbital-dependent. In a way,
the Ising transitions in BLG occur in part because of a change in the
\textquotedblleft internal structure\textquotedblright\ of the electron.

The discontinuous transitions between spin-polarized states at $\nu _{N}=2$
that occur when the bias is varied cause abrupt changes in the transport gap
that may explain some recent experimental results\cite{tutuc} on the
transport properties of the C2DEG in bilayer graphene. In the present work,
we study the behavior of the transport gap of the C2DEG with bias at odd
fillings of the higher Landau levels. We first show that the Ising behavior
found previously\cite{luo} should also occur in chirally-stacked multilayer
graphene, at least at zero bias. We then compute the energy of different
types of charged excitations of the Ising ground states. We consider bulk
quasiparticles (quasi-electrons and quasi-holes) and topological
quasiparticles (skyrmions and antiskyrmions). In each case, we evaluate the
transport gap due to the excitation of a particle-antiparticle pair using a
microscopic Hartree-Fock calculation. Our approach allows us to compute the
excitation gap at finite Zeeman and/or bias and takes into account the
possibility of intertwined spin and valley-pseudospin textures. When
appropriate (at zero bias or Zeeman coupling), we also compute the transport
gap using a long-wavelength approximation of the Hartree-Fock energy
functional. When the Zeeman coupling is taken into account, we find that
valley skyrmions are the lowest-energy excitations in bilayer graphene, at
least for $N=\pm 1$. This contrasts with previously studied Ising systems
where skyrmions are not the lowest-energy excitations\cite{lilliehook}. The
skyrmions that we find have either a spin or a valley-pseudospin texture,
but not both\cite{ezawa,gosh} and, as in monolayer graphene\cite{luo2}, they
exist only in a small range of bias or Zeeman coupling.

Magnetic Ising ferromagnets show an hysteretic behavior of the magnetization
as the magnetic field is swept upward or downward. In the absence of
disorder and at $T=0,$ the external magnetic field must exceed a certain
coercive field in order to produce the spin reversal. We find a similar
behavior with bias for the valley pseudospins and so we compute the coercive
bias for Landau levels $N=\pm 1.$ Owing to the fact that more valley-flip
transitions are possible in levels $N>0$ than in $N<0,$ we find that the
hysteretic behavior for $N>0$ is also more complex, with a very large
coercive field.

At finite temperature or in a disordered system, domains with particular
valley-pseudospin orientations are formed in the C2DEG. Regions of different
valley pseudospins are separated by domain walls which can have complicated
patterns. When domain walls are present, the electronic band structure and
so the electron-hole excitation gap is modulated in space. In the coherent
region of a domain wall, the valley pseudospins rotate from one polarization
to the other. For a linear domain wall, the coherent region forms a linear
channel of a few magnetic lengths in width. At the center of these regions,
the band gap is minimal. It should be possible, then, to excite
electron-holes pairs that are confined to a coherent region of a wall and
that have smaller excitation energy than the bulk electron-hole or the
skyrmion-antiskyrmion pairs. These mobile particles may be responsible\cite%
{jungwirth} for the increase in dissipation and breakdown of the quantum
Hall effect that is observed in some Ising QHFs\cite{poortere}. We compute
the energy of such excitations and find that, in some range of bias, they
can be the lowest-energy excitations.

This paper is organized in the following way. Sections II and III introduce
the tight-binding model and the electronic states in bilayer graphene as
well as most of the Hartree-Fock formalism necessary to derive the energy
gap for the particle-antiparticle pairs. Section IV discusses the hysteretic
behavior of the Ising QHFs and extend this behavior to other,
chirally-stacked, multilayer graphene structures. The transport gaps for the
electron-hole pairs, confined electron-hole pairs and skyrmion-antiskyrmion
pairs are computed in Secs. V,VI, and VII respectively. Section VIII is the
conclusion. The appendix provides a derivation of the projected
representation of a skyrmion that is used in Sec. VII.

\section{TIGHT-BINDING MODEL}

Bilayer graphene consists of two graphene layers separated by a distance $%
d=3.35$ \AA . In each layer, the crystal structure is a honeycomb lattice
that can be described as a triangular Bravais lattice with a basis of two
carbon atoms $A_{m},B_{m}$ where $m=1,2$ is the layer index. The in-plane
lattice constant is $a_{0}=2.46$ \AA . In the Bernal stacking, the upper $A$
sublattice is directly on top of the lower $B$ sublattice, while the upper $%
B $ sublattice is above the center of a hexagonal plaquette of the lower
layer. In the continuum approximation, the Hamiltonian is expanded around
the two non-equivalent valley points $K_{\pm }=\pm \frac{4\pi }{3}\left(
a_{0},0\right) $ of the Brillouin zone\cite{graphenereview}. In a transverse
magnetic field $\mathbf{B}=B_{0}\widehat{\mathbf{z}}$, the tight-binding
Hamiltonian for an electron with the valley index $\xi =\pm 1$ is given, in
the basis $\left\{ A_{1},B_{1},A_{2},B_{2}\right\} ,$ by%
\begin{equation}
H_{\pm }^{\left( 0\right) }=\left( 
\begin{array}{cccc}
\delta _{0}+\frac{\Delta _{B}}{2} & \pm \alpha _{0}a^{\pm } & \pm \alpha
_{4}a^{\mp } & -\gamma _{1} \\ 
\pm \alpha _{0}a^{\mp } & \frac{\Delta _{B}}{2} & 0 & \pm \alpha _{4}a^{\mp }
\\ 
\pm \alpha _{4}a^{\pm } & 0 & -\frac{\Delta _{B}}{2} & \pm \alpha _{0}a^{\pm
} \\ 
-\gamma _{1} & \pm \alpha _{4}a^{\pm } & \pm \alpha _{0}a^{\mp } & \delta
_{0}-\frac{\Delta _{B}}{2}%
\end{array}%
\right) ,
\end{equation}%
where $\alpha _{i}=\sqrt{\frac{3}{2}}\frac{a_{0}}{\ell }\gamma _{i}$ with $%
\ell =\sqrt{\hslash c/eB}$ the magnetic length and $\gamma _{j}$ the hopping
parameters. A recent calculation\cite{jung} gives $\gamma _{0}=2.61$ eV for
the nearest-neighbor hopping, $\gamma _{1}=-0.361$ eV for the interlayer
hopping between carbon atoms that are immediately above one another (i.e. $%
A_{1}-B_{2}$) and $\gamma _{4}=-0.138$ eV for the interlayer next
nearest-neighbor hopping between carbons atoms in the same sublattice (i.e. $%
A_{1}-A_{2}$ and $B_{1}-B_{2}$). The magnetic field considered in the
present work is assumed sufficiently large for the warping term $\gamma _{3}$
to be negligible\cite{mccann,manuel}. The parameter $\delta _{0}=0.015$ eV
represents the difference in the crystal field between sites $A_{1},B_{2}$
and $A_{2},B_{1}$. The ladder operators $a^{-}=a$ and $a^{+}=a^{\dag }$ are
defined in the Landau gauge $\mathbf{A}=\left( 0,Bx,0\right) .$

In the absence of a magnetic field, the electronic dispersion has four bands
which we denote by $j=1,2,3,4$ with $j=1,2$ the negative and $j=3,4$ the
positive energy bands. The two middle bands $j=2,3$ touch each other at the
six valley points $K_{\pm }$ where they have zero energy in the absence of
bias. The high-energy bands $j=1,4$ are separated from the middle two energy
bands by a gap of order $\gamma _{1}.$ In a quantizing magnetic field, each
band is split into Landau levels. We shall be concerned only with bands $%
j=2,3$ in this work. The eigenstates of $H_{\pm }^{\left( 0\right) }$ for
the levels of these bands have energies $E_{\xi ,n,j}^{\left( 0\right) }$
with $n=0,1,2,...$ and the corresponding eigenvectors are given by

\begin{equation}
\psi _{\xi =-1,n,j,X}\left( \mathbf{r},z\right) =\left( 
\begin{array}{c}
b_{\xi ,n,j,1}h_{n-1,X}\left( \mathbf{r}\right) \chi _{1}\left( z\right) \\ 
b_{\xi ,n,j,2}h_{n,X}\left( \mathbf{r}\right) \chi _{1}\left( z\right) \\ 
b_{\xi ,n,j,3}h_{n-2,X}\left( \mathbf{r}\right) \chi _{2}\left( z\right) \\ 
b_{\xi ,n,j,4}h_{n-1,X}\left( \mathbf{r}\right) \chi _{2}\left( z\right)%
\end{array}%
\right) ,  \label{spinor1}
\end{equation}%
and%
\begin{equation}
\psi _{\xi =+1,n,j,X}\left( \mathbf{r},z\right) =\left( 
\begin{array}{c}
b_{\xi ,n,j,1}h_{n-1,X}\left( \mathbf{r}\right) \chi _{1}\left( z\right) \\ 
b_{\xi ,n,j,2}h_{n-2,X}\left( \mathbf{r}\right) \chi _{1}\left( z\right) \\ 
b_{\xi ,n,j,3}h_{n,X}\left( \mathbf{r}\right) \chi _{2}\left( z\right) \\ 
b_{\xi ,n,j,4}h_{n-1,X}\left( \mathbf{r}\right) \chi _{2}\left( z\right)%
\end{array}%
\right) ,  \label{spinor2}
\end{equation}%
where $X$ is the guiding-center index. The functions $\left\vert \chi
_{i}\left( z\right) \right\vert ^{2}=\delta \left( z-z_{i}\right) ,$ where $%
z_{i}$ is the position of the graphene layer $i.$ By definition, $%
h_{n,X}\left( \mathbf{r}\right) =0$ if $n<0,$ otherwise%
\begin{equation}
h_{n,X}\left( \mathbf{r}\right) =\frac{1}{\sqrt{L_{y}}}e^{-iXy/\ell
^{2}}\varphi _{n}\left( x-X\right) ,
\end{equation}%
where $n$ is the level index and $L_{y}$ the sample dimension in the $y$
direction. The functions $\varphi _{n}\left( x\right) $ are the eigenstates
of the one-dimensional harmonic oscillator.

In a minimal tight-binding model where only $\gamma _{0},\gamma _{1}$ are
kept, levels $n=0,1$ are degenerate and have zero energy. When spin and
valley degrees of freedom are counted and the Zeeman term is neglected,
there are in total eight degenerate levels with zero energy. This octet of
states defines the Landau level $N=0.$ The higher Landau levels have $n\geq
2 $ with $j=2,3.$ The positive energy Landau levels $N=n-1$ have band index $%
j=3$ while the negative energy ones have $N=-\left( \left\vert n\right\vert
-1\right) $ with $j=2.$ This notation is illustrated in Fig. 1 of Ref. %
\onlinecite{luo}. Contrary to $N=0,$ the higher-energy Landau levels $%
\left\vert N\right\vert >0$ are fourfold degenerate. Each level with quantum
numbers $\left( \xi ,\sigma ,n,j\right) $ has the macroscopic degeneracy $%
N_{\varphi }=S/2\pi \ell ^{2},$ where $S$ is the C2DEG area, associated with
the guiding-center index $X$.

In this work, we assume that Landau level mixing can be neglected. As shown
in Ref. \onlinecite{manuel}, this limits the electric bias to $\Delta _{B}$ $%
\lesssim 0.1$ eV for $B=10$ T and $\Delta _{B}$ $\lesssim 0.2$ eV for $B=30$
T. We project the Hamiltonian into level $N$ so that the electron field
operator is given by%
\begin{equation}
\Psi _{\mu ,N}\left( \mathbf{r},z\right) =\sum_{X}\psi _{\mu ,N,X}\left( 
\mathbf{r},z\right) c_{\mu ,N,X},  \label{field}
\end{equation}%
where $c_{\mu ,N,X}$ annihilates an electron is state $\left( \mu
,N,X\right) $ and the super-index $\mu =\left( \xi ,\sigma \right) $
corresponds to the four states%
\begin{eqnarray}
\mu &=&1\rightarrow \left( K_{+},+\right) , \\
\mu &=&2\rightarrow \left( K_{+},-\right) , \\
\mu &=&3\rightarrow \left( K_{-},+\right) , \\
\mu &=&4\rightarrow \left( K_{-},-\right) ,
\end{eqnarray}%
where the second index in the parenthesis is for the spin. Hereafter, we use
the notation $\mu _{\xi }=\xi $ and $\mu _{\sigma }=\sigma .$

\section{HARTREE-FOCK FORMALISM FOR THE\ HIGHER LANDAU LEVELS}

The various energy gaps are computed in this work in the Hartree-Fock
approximation (HFA). In a recent work on the Ising QHFs\cite{luo}, the HFA
was used to compute the energy of ground states with uniform spin and/or
valley polarizations. In this section, the HFA is extended to the study of
excitations with spin and/or pseudospin textures. To simplify the notation,
the subscript $N=\left( n,j\right) $ is omitted whenever possible, since the
Hilbert space is restricted to level $N$ only.

\subsection{Hartree-Fock Hamiltonian}

The Hartree-Fock Hamiltonian for electrons in level $N$ is given by

\begin{eqnarray}
\frac{H_{HF}}{N_{\varphi }} &=&-\alpha _{d}\frac{\nu _{N}^{2}}{4}+\frac{1}{2}%
\alpha _{d}\nu _{N}\left( \rho _{1}\left( 0\right) +\rho _{2}\left( 0\right)
\right)  \label{hfenergy} \\
&&+\sum_{\mu }\left( E_{\mu _{\xi }}^{\left( 0\right) }-\frac{1}{2}\mu
_{\sigma }\Delta _{Z}\right) \rho _{\mu ,\mu }\left( 0\right)  \notag \\
&&-\alpha _{d}\left[ \left\langle \rho _{1}\left( 0\right) \right\rangle
\rho _{2}\left( 0\right) +\left\langle \rho _{2}\left( 0\right)
\right\rangle \rho _{1}\left( 0\right) \right]  \notag \\
&&+\sum_{\mu ,\mu ^{\prime }}\overline{\sum_{\mathbf{q}}}H^{\left( \mu _{\xi
},\mu _{\xi }^{\prime }\right) }\left( \mathbf{q}\right) \left\langle \rho
_{\mu ,\mu }\left( -\mathbf{q}\right) \right\rangle \rho _{\mu ^{\prime
},\mu ^{\prime }}\left( \mathbf{q}\right)  \notag \\
&&-\sum_{\mu ,\mu ^{\prime }}\sum_{\mathbf{q}}X^{\left( \mu _{\xi },\mu
_{\xi }^{\prime }\right) }\left( \mathbf{q}\right) \left\langle \rho _{\mu
,\mu ^{\prime }}\left( -\mathbf{q}\right) \right\rangle \rho _{\mu ^{\prime
},\mu }\left( \mathbf{q}\right) ,  \notag
\end{eqnarray}%
where $\Delta _{Z}=g\mu _{B}B,$ with $g=2$ and $\mu _{B}$ the Bohr magneton,
is the Zeeman coupling. The bar over the summation in the Hartree term means
that the $\mathbf{q}=0$ is omitted from the sum because a positive
background is considered in order to make the system neutral. We have
defined the constant 
\begin{equation}
\alpha _{d}=\frac{d}{\ell }\left( \frac{e^{2}}{\kappa \ell }\right)
\end{equation}%
and $\nu _{N}=N_{e}/N_{\varphi }$ is the filling factor of level $N$ with $%
N_{e}$ the number of electrons in this level. In $\alpha _{d},$ $\kappa $ is
the dielectric constant of the substrate. In all the calculations of this
paper, $\kappa =2.5.$

The ground state average value of the operators 
\begin{eqnarray}
\rho _{\mu ,\mu ^{\prime }}\left( \mathbf{q}\right) &=&\frac{1}{N_{\varphi }}%
\sum_{X,X^{\prime }}e^{-\frac{i}{2}q_{x}\left( X+X^{\prime }\right) }
\label{rho} \\
&&\times \delta _{X,X^{\prime }+q_{y}\ell ^{2}}c_{\mu ,N}^{\dagger }c_{\mu
^{\prime },X^{\prime }}  \notag
\end{eqnarray}%
can be considered as the order parameters of a specific phase of the
electron gas. The operators 
\begin{equation}
\rho _{i}\left( 0\right) =\sum_{\mu }n_{i,\mu _{\xi }}\rho _{\mu ,\mu
}\left( \mathbf{q}=0\right) ,
\end{equation}%
with $i=1,2$ and the projectors%
\begin{eqnarray}
n_{1,+} &=&\left\vert b_{4,+}\right\vert ^{2}+\left\vert b_{3,+}\right\vert
^{2},  \label{yy0} \\
n_{1,-} &=&\left\vert b_{1,-}\right\vert ^{2}+\left\vert b_{2,-}\right\vert
^{2}, \\
n_{2,+} &=&\left\vert b_{1,+}\right\vert ^{2}+\left\vert b_{2,+}\right\vert
^{2}, \\
n_{2,-} &=&\left\vert b_{3,-}\right\vert ^{2}+\left\vert b_{4,-}\right\vert
^{2},  \label{yy5}
\end{eqnarray}%
and%
\begin{eqnarray}
n_{1} &=&n_{1,+}+n_{1,-}, \\
n_{2} &=&n_{2,+}+n_{2,-},
\end{eqnarray}%
are defined such that $\left\langle \rho _{i}\left( 0\right) \right\rangle $
is the filling factor of \textit{layers} $i=1,2$. It is important to
remember that, unlike level $N=0,$ layer and valley indices are not
equivalent in the higher Landau levels.

The Hartree and Fock interactions that enter $H_{HF}$ are defined by (here, $%
x=q\ell \,$)%
\begin{eqnarray}
H^{\left( \xi ,\xi \right) }\left( x\right) &=&\left( \frac{e^{2}}{\kappa
\ell }\right) \frac{V^{\left( \xi ,\xi \right) }\left( x\right) }{x},
\label{yy1} \\
H^{\left( \xi ,\overline{\xi }\right) }\left( x\right) &=&\left( \frac{e^{2}%
}{\kappa \ell }\right) \frac{V^{\left( \xi ,\overline{\xi }\right) }\left(
x\right) }{x}, \\
X^{\left( \xi ,\xi \right) }\left( x\right) &=&\left( \frac{e^{2}}{\kappa
\ell }\right) \int_{0}^{\infty }dyV^{\left( \xi ,\xi \right) }\left(
y\right) J_{0}\left( xy\right) , \\
X^{\left( \xi ,\overline{\xi }\right) }\left( x\right) &=&\left( \frac{e^{2}%
}{\kappa \ell }\right) \int_{0}^{\infty }dyV^{\left( \xi ,\overline{\xi }%
\right) }\left( y\right) J_{0}\left( xy\right) ,
\end{eqnarray}%
with $\overline{\xi }=-\xi $ and ($G_{i}^{\left( \xi \right) }=G_{i}^{\left(
\xi \right) }\left( x\right) ,$ for $i=1,2$) and 
\begin{eqnarray}
V^{\left( \xi ,\xi \right) }\left( x\right) &=&G_{1}^{\left( \xi \right)
}G_{1}^{\left( \xi \right) }+G_{2}^{\left( \xi \right) }G_{2}^{\left( \xi
\right) } \\
&&+e^{-qd}\left[ G_{1}^{\left( \xi \right) }G_{2}^{\left( \xi \right)
}+G_{2}^{\left( \xi \right) }G_{1}^{\left( \xi \right) }\right] ,  \notag \\
V^{\left( \xi ,\overline{\xi }\right) }\left( x\right) &=&G_{1}^{\left( \xi
\right) }G_{2}^{\left( \overline{\xi }\right) }+G_{2}^{\left( \xi \right)
}G_{1}^{\left( \overline{\xi }\right) } \\
&&+e^{-qd}\left[ G_{1}^{\left( \xi \right) }G_{1}^{\left( \overline{\xi }%
\right) }+G_{2}^{\left( \xi \right) }G_{2}^{\left( \overline{\xi }\right) }%
\right] ,  \notag
\end{eqnarray}%
with the functions%
\begin{eqnarray}
G_{1}^{\left( \xi \right) }\left( x\right) &=&e^{-x^{2}/4}\left[ \left\vert
b_{1,\xi }\right\vert ^{2}L_{n-1}\left( \frac{x^{2}}{2}\right) \right. \\
&&\left. +\left\vert b_{2,\xi }\right\vert ^{2}L_{n}\left( \frac{x^{2}}{2}%
\right) \right]  \notag \\
G_{2}^{\left( \xi \right) }\left( x\right) &=&e^{-x^{2}/4}\left[ \left\vert
b_{3,\xi }\right\vert ^{2}L_{n-2}\left( \frac{x^{2}}{2}\right) \right.
\label{yy3} \\
&&+\left. \left\vert b_{4,\xi }\right\vert ^{2}L_{n-1}\left( \frac{x^{2}}{2}%
\right) \right] ,  \notag
\end{eqnarray}%
where $L_{n}\left( x\right) $ is a Laguerre polynomial.

The Hartree-Fock energy per electron is 
\begin{eqnarray}
\frac{E_{HF}}{N_{e}} &=&\frac{1}{\nu _{N}}\alpha _{d}\left[ \left\langle
\rho _{1}\left( 0\right) \right\rangle -\left\langle \rho _{2}\left(
0\right) \right\rangle \right] ^{2}  \label{ener1} \\
&&+\frac{1}{\nu _{N}}\sum_{\mu }\left( E_{\mu _{\xi }}^{\left( 0\right) }-%
\frac{\mu _{\sigma }\Delta _{Z}}{2}\right) \left\langle \rho _{\mu ,\mu
}\left( 0\right) \right\rangle  \notag \\
&&+\frac{1}{2\nu _{N}}\sum_{\mu ,\mu ^{\prime }}\overline{\sum_{\mathbf{q}}}%
H^{\left( \xi ,\zeta \right) }\left( \mathbf{q}\right) \left\langle \rho
_{\mu ,\mu }\left( -\mathbf{q}\right) \right\rangle \left\langle \rho _{\mu
^{\prime },\mu ^{\prime }}\left( \mathbf{q}\right) \right\rangle  \notag \\
&&-\frac{1}{2\nu _{N}}\sum_{\mu ,\mu ^{\prime }}\sum_{\mathbf{q}}X^{\left(
\xi ,\zeta \right) }\left( \mathbf{q}\right) \left\langle \rho _{\mu ,\mu
^{\prime }}\left( -\mathbf{q}\right) \right\rangle \left\langle \rho _{\mu
^{\prime },\mu }\left( \mathbf{q}\right) \right\rangle .  \notag
\end{eqnarray}

\subsection{Order parameters and single-particle Green's functions}

To compute the order parameters, we define the matrix of Matsubara Green's
functions 
\begin{equation}
G_{\mu ,\mu ^{\prime }}\left( X,X^{\prime },\tau \right) =-\left\langle
T_{\tau }c_{\mu ,X}\left( \tau \right) c_{\mu ^{\prime },X^{\prime
}}^{\dagger }\left( 0\right) \right\rangle ,  \label{gf1}
\end{equation}%
where $T_{\tau }$ is the time-ordering operator. Its Fourier transform is
defined by 
\begin{eqnarray}
G_{\mu ,\mu ^{\prime }}\left( \mathbf{q,}\tau \right) &=&\frac{1}{N_{\varphi
}}\sum_{X,X^{\prime }}e^{-\frac{i}{2}q_{x}\left( X+X^{\prime }\right) }
\label{GX} \\
&&\times \delta _{X,X^{\prime }-q_{y}\ell ^{2}}G_{\mu ,\mu ^{\prime }}\left(
X,X^{\prime },\tau \right) ,  \notag
\end{eqnarray}%
so that%
\begin{equation}
\left\langle \rho _{\mu ^{\prime },\mu }\left( \mathbf{q}\right)
\right\rangle =G_{\mu ,\mu ^{\prime }}\left( \mathbf{q,}\tau =0^{-}\right) .
\end{equation}

The matrix of Green's functions is computed from its Hartree-Fock equation
of motion 
\begin{eqnarray}
&&\left[ \hslash i\omega _{n}-\left( E_{\mu }-\widetilde{\mu }\right) \right]
G_{\mu ,\mu ^{\prime }}\left( \mathbf{q},\omega _{n}\right) =\hslash \delta
_{\mathbf{q},0}\delta _{\mu ,\mu ^{\prime }}  \label{eqmotion} \\
&&+\sum_{\mu ^{\prime \prime }}\overline{\sum_{\mathbf{q}^{\prime }\neq 
\mathbf{q}}}H^{\left( \mu _{\xi }^{\prime \prime },\mu _{\xi }\right)
}\left( \mathbf{q}^{\prime }-\mathbf{q}\right) \left\langle \rho _{\mu
^{\prime \prime },\mu ^{\prime \prime }}\left( \mathbf{q-q}^{\prime }\right)
\right\rangle  \notag \\
&&\times e^{-i\left( \mathbf{q}\times \mathbf{q}^{\prime }\right) \cdot 
\widehat{\mathbf{z}}\ell ^{2}/2}G_{\mu ,\mu ^{\prime }}\left( \mathbf{q}%
^{\prime },\omega _{n}\right)  \notag \\
&&-\sum_{\mu ^{\prime \prime }}\sum_{\mathbf{q}^{\prime }}X^{\left( \mu
_{\xi }^{\prime \prime },\mu _{\xi }\right) }\left( \mathbf{q}^{\prime }-%
\mathbf{q}\right) \left\langle \rho _{\mu ^{\prime \prime },\mu }\left( 
\mathbf{q-q}^{\prime }\right) \right\rangle  \notag \\
&&\times e^{-i\left( \mathbf{q}\times \mathbf{q}^{\prime }\right) \cdot 
\widehat{\mathbf{z}}\ell ^{2}/2}G_{\mu ^{\prime \prime },\mu ^{\prime
}}\left( \mathbf{q}^{\prime },\omega _{n}\right) ,  \notag
\end{eqnarray}%
where $\widetilde{\mu }$ is the chemical potential and the renormalized
energies are given by 
\begin{equation}
E_{\mu }=E_{\mu _{\xi }}^{\left( 0\right) }-\frac{1}{2}\mu _{\sigma }\Delta
_{Z}+\sum_{\mu ^{\prime }}A^{\left( \mu _{\xi },\mu _{\xi }^{\prime }\right)
}\left\langle \rho _{\mu ^{\prime },\mu ^{\prime }}\left( 0\right)
\right\rangle ,
\end{equation}%
where%
\begin{eqnarray}
A^{\left( \pm ,\pm \right) } &=&-2\alpha _{d}n_{2,\pm }n_{1,\pm }, \\
A^{\left( \pm ,\mp \right) } &=&-\alpha _{d}\left(
n_{2,+}n_{1,-}+n_{1,+}n_{2,-}\right) .
\end{eqnarray}

\subsection{Hartree-Fock energy in the pseudospin language}

It is instructive to rewrite the Hartree-Fock energy in a pseudospin
language. Since the skyrmions found in the HFA do not contain intertwined
spin and valley-pseudospin textures, it is sufficient to define pure spin
and valley-pseudospin fields. For the total spin field, including both
valleys, 
\begin{eqnarray}
S_{z} &=&\frac{1}{2}\sum_{\mu }\mu _{\sigma }\left\langle \rho _{\mu ,\mu
}\right\rangle , \\
S_{x}+iS_{y} &=&\left\langle \rho _{1,2}\right\rangle +\left\langle \rho
_{3,4}\right\rangle .
\end{eqnarray}%
For the total valley pseudospin, including both spin states, 
\begin{eqnarray}
P_{z} &=&\frac{1}{2}\sum_{\mu }\mu _{\xi }\left\langle \rho _{\mu ,\mu
}\right\rangle , \\
P_{x}+iP_{y} &=&\left\langle \rho _{1,3}\right\rangle +\left\langle \rho
_{2,4}\right\rangle .
\end{eqnarray}%
Finally, for the total electronic density%
\begin{equation}
\rho =\sum_{\mu }\left\langle \rho _{\mu ,\mu }\right\rangle .
\end{equation}%
The Fourier transform of these fields is defined by%
\begin{equation}
\mathbf{F}\left( \mathbf{r}\right) =\frac{1}{S}\sum_{\mathbf{q}}e^{i\mathbf{q%
}\cdot \mathbf{r}}\mathbf{f}\left( \mathbf{q}\right) ,
\end{equation}%
where $\mathbf{F}\left( \mathbf{r}\right) $ is one of the fields and $%
\mathbf{f}\left( \mathbf{q}\right) $ is defined in term of the order
parameters $\left\{ \left\langle \rho _{\mu ,\mu ^{\prime }}\left( \mathbf{q}%
\right) \right\rangle \right\} .$ It does not include the form factors that
reflect the character of the different orbitals present in the electronic
spinor (see Eqs. (\ref{spinor1}) and (\ref{spinor2}))$.$ We say that $%
\mathbf{F}\left( \mathbf{r}\right) $ is given in the guiding-center
representation (GCR).

We define the field $\mathbf{S}_{\xi }$ as the part of the total spin field
which is in valley $\xi $ while $\mathbf{P}_{\bot ,\sigma }$ is the part of
the total in-plane valley-pseudospin field with spin $\sigma $. The fields
are coupled by the constraint%
\begin{eqnarray}
&&\sum_{\mathbf{q}}\left[ \frac{1}{4}\left\vert \rho \right\vert
^{2}+\left\vert P_{z}\right\vert ^{2}+2\left\vert P_{\bot ,+}\right\vert
^{2}+2\left\vert P_{\bot ,-}\right\vert ^{2}\right.  \label{sumrule3} \\
&&\left. +2\left\vert \mathbf{S}_{+}\right\vert ^{2}+2\left\vert \mathbf{S}%
_{-}\right\vert ^{2}+2\left\vert \left\langle \rho _{1,4}\right\rangle
\right\vert ^{2}+2\left\vert \left\langle \rho _{2,3}\right\rangle
\right\vert ^{2}\right]  \notag \\
&=&\nu _{N},  \notag
\end{eqnarray}%
where $\rho =\rho \left( \mathbf{q}\right) ,P_{z}=P_{z}\left( \mathbf{q}%
\right) ,$ etc.

In the GCR, the Hartree-Fock energy per electron $E_{HF}$ becomes

\begin{eqnarray}
\frac{E_{HF}}{N_{e}} &=&\frac{1}{\nu _{N}}\sum_{\xi }\left( \frac{\rho
\left( 0\right) }{2}E_{\mu _{\xi }}^{\left( 0\right) }+\xi E_{\mu _{\xi
}}^{\left( 0\right) }P_{z}\left( 0\right) \right)  \label{totalener} \\
&&+\frac{1}{4\nu _{N}}\alpha _{d}\left( \left\langle \rho _{1}\left(
0\right) \right\rangle -\left\langle \rho _{2}\left( 0\right) \right\rangle
\right) ^{2}-\Delta _{Z}S_{z}\left( 0\right)  \notag \\
&&+\frac{1}{8\nu _{N}}\sum_{\mathbf{q}}\Lambda _{\rho ,\rho }\left( \mathbf{q%
}\right) \left\vert \rho \left( \mathbf{q}\right) \right\vert ^{2}  \notag \\
&&+\frac{1}{2\nu _{N}}\sum_{\mathbf{q}}\Lambda _{z,z}\left( \mathbf{q}%
\right) \left\vert P_{z}\left( \mathbf{q}\right) \right\vert ^{2}  \notag \\
&&+\frac{1}{2\nu _{N}}\sum_{\mathbf{q}}\Lambda _{\rho ,z}\left( \mathbf{q}%
\right) \rho \left( -\mathbf{q}\right) P_{z}\left( \mathbf{q}\right)  \notag
\\
&&-\frac{1}{\nu _{N}}\sum_{\mathbf{q}}\sum_{\xi }X^{\left( \xi ,\xi \right)
}\left( \mathbf{q}\right) \left\vert \mathbf{S}_{\xi }\left( \mathbf{q}%
\right) \right\vert ^{2}  \notag \\
&&-\frac{1}{\nu _{N}}\sum_{\mathbf{q}}\sum_{\sigma }X^{\left( +,-\right)
}\left( \mathbf{q}\right) \left\vert \mathbf{P}_{\bot ,\sigma }\left( 
\mathbf{q}\right) \right\vert ^{2}  \notag \\
&&-\frac{1}{2\nu _{N}}\sum_{\mathbf{q}}X^{\left( +,-\right) }\left( \mathbf{q%
}\right) \left\vert \left\langle \rho _{1,4}\left( \mathbf{q}\right)
\right\rangle \right\vert ^{2}  \notag \\
&&-\frac{1}{2\nu _{N}}\sum_{\mathbf{q}}X^{\left( +,-\right) }\left( \mathbf{q%
}\right) \left\vert \left\langle \rho _{2,3}\left( \mathbf{q}\right)
\right\rangle \right\vert ^{2}.  \notag
\end{eqnarray}%
The interactions are functions of $x=\left\vert \mathbf{q\ell }\right\vert $
only and are given by%
\begin{eqnarray}
\Lambda _{\rho ,\rho } &=&\sum_{\xi }\left[ H^{\left( \xi ,\xi \right)
}+H^{\left( \xi ,\overline{\xi }\right) }-\frac{1}{2}X^{\left( \xi ,\xi
\right) }\right] , \\
\Lambda _{\rho ,z} &=&\sum_{\xi }\xi \left[ H^{\left( \xi ,\xi \right) }-%
\frac{1}{2}X^{\left( \xi ,\xi \right) }\right] , \\
\Lambda _{z,z} &=&\sum_{\xi }\left[ H^{\left( \xi ,\xi \right) }-H^{\left(
\xi ,\overline{\xi }\right) }-\frac{1}{2}X^{\left( \xi ,\xi \right) }\right]
.
\end{eqnarray}

\subsection{Gradient approximation of the Hartree-Fock energy}

In the limit of zero Zeeman or bias coupling, skyrmions usually become very
large and their energies cannot be computed by the microscopic HFA of Eq. (%
\ref{eqmotion}) since the matrix of Green's functions becomes very large. In
this limit, it is more useful to make a gradient approximation of the total
energy in Eq. (\ref{totalener}), taking advantage of the fact that the spin
or pseudospin texture varies slowly in space\cite{moon} so that only the
first few terms in the expansion can be kept. In levels $\left\vert
N\right\vert >0$ of bilayer graphene, the gradient approximation works well
for spin skyrmion but, as we will show below, cannot be applied to
valley-pseudospin skyrmions since these later topological excitations have a
small finite size at zero bias.

\subsubsection{Spin textures}

When all electrons occupy the valley $K_{\xi }$ at $\nu _{N}=1$, there is no
intervalley coherence and Eq. (\ref{totalener}) at zero bias and zero Zeeman
coupling simplifies to

\begin{eqnarray}
E_{\mathrm{spin}} &=&N_{\varphi }E_{\xi }^{\left( 0\right) }+\frac{1}{4}%
N_{\varphi }\alpha _{d}\left( n_{1,\xi }-n_{2,\xi }\right) ^{2}
\label{espin} \\
&&+\frac{1}{2}N_{\varphi }\sum_{\mathbf{q}}\widetilde{\Lambda }_{\rho ,\rho
}^{\left( \xi \right) }\left( \mathbf{q}\right) \left\vert \rho \left( 
\mathbf{q}\right) \right\vert ^{2}  \notag \\
&&-N_{\varphi }\sum_{\mathbf{q}}X^{\left( \xi ,\xi \right) }\left( \mathbf{q}%
\right) \left\vert \mathbf{S}_{\xi }\left( \mathbf{q}\right) \right\vert
^{2},  \notag
\end{eqnarray}%
with%
\begin{equation}
\widetilde{\Lambda }_{\rho ,\rho }^{\left( \xi \right) }=H^{\left( \xi ,\xi
\right) }-\frac{1}{2}X^{\left( \xi ,\xi \right) }.
\end{equation}%
Note that, at zero bias,%
\begin{eqnarray}
n_{1,\pm } &=&n_{2,\mp },  \label{r1} \\
E_{+}^{\left( 0\right) } &=&E_{-}^{\left( 0\right) }\equiv E^{\left(
0\right) }, \\
H^{\left( +,+\right) } &=&H^{\left( -,-\right) }\equiv H, \\
X^{\left( +,+\right) } &=&X^{\left( -,-\right) }\equiv X, \\
V^{\left( +,+\right) } &=&V^{\left( -,-\right) }\equiv V,  \label{r5}
\end{eqnarray}%
so that $E_{\mathrm{spin}}$ is independent of the valley index.

For a spin-textured excitation, all electrons remain in the same valley but
some of them flip their spin. If we neglect the charge density term (i.e.
the third term on the right-hand side of Eq. (\ref{espin})), then a gradient
approximation gives for the excitation energy with respect to the ground
state 
\begin{equation}
\delta E_{\mathrm{spin}}=\frac{1}{2}\rho _{S}\sum_{\nu }\int d\mathbf{r}%
\left( \partial _{\nu }\mathbf{s}_{\xi }\left( \mathbf{r}\right) \right)
^{2},  \label{efunspin}
\end{equation}%
where $\nu =x,y$ and the field $\mathbf{s}_{\xi }\left( \mathbf{r}\right)
\equiv 2S\mathbf{S}_{\xi }\left( \mathbf{r}\right) $ (where $S$ is the 2DEG
area) has unit modulus\cite{note4}. The spin stiffness is defined by%
\begin{equation}
\rho _{S}=\frac{1}{16\pi }\left( \frac{e^{2}}{\kappa \ell }\right)
\int_{0}^{\infty }dyV\left( y\right) y^{2}.  \label{stifspin}
\end{equation}%
Eq. (\ref{efunspin}), with the condition that $\left\vert \mathbf{s}_{\xi
}\left( \mathbf{r}\right) \right\vert =1,$ is the nonlinear $\sigma $ model
(NL$\sigma $M). Note that, $\delta E_{\mathrm{spin}}$ is the energy of an
uncharged spin texture. No electron is added or removed to the system. To
compute the energy of a charged spin texture, it is necessary to consider
the chemical potential\cite{moon}. There is no need to worry about this
problem in the present work since we are interested only in the transport
gap which is the energy required to excite a skyrmion-antiskyrmion pair.
This energy can be expressed solely\cite{girvin} in terms of $\rho _{S}$
(see Eq. (\ref{eskask}) below).

A finite Zeeman coupling favors small spin textures while the charge-density
term favors large textures. At finite Zeeman coupling, the size of the spin
texture is a compromise between these two terms and it can be small. The
gradient approximation is then no longer valid. A microscopic Hartree-Fock
calculation must be done to compute the energy of small spin textures.

\subsubsection{Valley-pseudospin textures}

When all electrons occupy the spin state $\sigma =+1$ at $\nu _{N}=1$, there
is no spin coherence and the total Hartree-Fock energy at zero bias is

\begin{eqnarray}
E_{\mathrm{valley}} &=&N_{\varphi }\left( E^{\left( 0\right) }-\frac{\Delta
_{B}}{2}\right)  \label{H1} \\
&&+\frac{1}{16}N_{\varphi }\alpha _{d}\left( n_{1}-n_{2}\right) ^{2}  \notag
\\
&&+N_{\varphi }\alpha _{d}\left( n_{1,+}-n_{1,-}\right) ^{2}P_{z}^{2}(0) 
\notag \\
&&+\frac{1}{4}N_{\varphi }\sum_{\mathbf{q}}\Upsilon _{\rho ,\rho }\left( 
\mathbf{q}\right) \left\vert \rho \left( \mathbf{q}\right) \right\vert ^{2} 
\notag \\
&&+N_{\varphi }\sum_{\mathbf{q}}\Upsilon _{z,z}\left( \mathbf{q}\right)
\left\vert P_{z,+}\left( \mathbf{q}\right) \right\vert ^{2}  \notag \\
&&-N_{\varphi }\sum_{\mathbf{q}}X^{\left( +,-\right) }\left( \mathbf{q}%
\right) \left\vert \mathbf{P}_{\bot ,+}\left( \mathbf{q}\right) \right\vert
^{2},  \notag
\end{eqnarray}%
where%
\begin{eqnarray}
\Upsilon _{\rho ,\rho } &=&H-X+H^{\left( +,-\right) }, \\
\Upsilon _{z,z} &=&H-X-H^{\left( +,-\right) }.
\end{eqnarray}

In a valley-pseudospin texture, no spin are flipped. Neglecting again the
charge-density term, the gradient approximation gives for the excitation
energy with respect to an Ising ground state with $P_{z}\left( 0\right) =\xi
/2$ \textbf{\ }%
\begin{eqnarray}
\delta E_{\mathrm{valley}} &=&K\int d\mathbf{r}\left[ p_{z}^{2}\left( 
\mathbf{r}\right) -1\right]  \label{evallee} \\
&&+\frac{1}{2}\sum_{\nu }\int d\mathbf{r}\left[ \rho _{z}\left( \partial
_{\nu }p_{z}\left( \mathbf{r}\right) \right) ^{2}+\rho _{\bot }\left(
\partial _{\nu }p_{\bot }\left( \mathbf{r}\right) \right) ^{2}\right]  \notag
\\
&&+N_{\varphi }\alpha _{d}\left( n_{1,+}-n_{1,-}\right) ^{2}\left[ \left(
P_{z}(q=0)\right) ^{2}-\frac{1}{4}\right] ,  \notag
\end{eqnarray}%
where $\mathbf{p}\left( \mathbf{r}\right) =2S\mathbf{P}\left( \mathbf{r}%
\right) $ is a unit field and the pseudospin stiffnesses are given by%
\begin{equation}
\rho _{\bot }=\frac{1}{16\pi }\left( \frac{e^{2}}{\kappa \ell }\right)
\int_{0}^{\infty }dyV^{\left( +,-\right) }\left( y\right) y^{2},
\label{stif1}
\end{equation}%
and%
\begin{equation}
\rho _{z}=\rho _{S}=\frac{1}{16\pi }\left( \frac{e^{2}}{\kappa \ell }\right)
\int_{0}^{\infty }dyV\left( y\right) y^{2},  \label{stif2}
\end{equation}%
\newline
while the easy-axis anisotropy term is given by

\begin{equation}
K=-\frac{1}{8\pi \ell ^{2}}\left( X\left( 0\right) -X^{+,-}\left( 0\right)
\right) .  \label{stif3}
\end{equation}%
The last term in Eq. (\ref{evallee}) is related to the capacitive energy. If 
$N_{\overline{\xi }}<<N_{\varphi },$ it can be rewritten as%
\begin{equation}
-\alpha _{d}\left( n_{1,R}-n_{1,L}\right) ^{2}\frac{N_{\xi }N_{\overline{\xi 
}}}{N_{\varphi }}\approx -\alpha _{d}\left( n_{1,R}-n_{1,L}\right) ^{2}N_{%
\overline{\xi }},
\end{equation}%
where $N_{\xi }$ is the number of electrons in valley $\xi .$ Even though
the bias is zero, there is a capacitive energy because in each state $K_{\xi
},$ there is a charge imbalance between the two layers which is given by 
\begin{equation}
\rho _{1}\left( 0\right) -\rho _{2}\left( 0\right) =\xi \left(
n_{1,+}-n_{2,+}\right) .
\end{equation}

Numerically, $\rho _{S},\rho _{\bot }$ and $K$ are all positive quantities.
As an example, for $N=-1,B=10$ T$,$ their value is $\rho _{z}=0.97$ meV$%
,\rho _{\bot }=0.85$ meV, $K=-0.048$ meV/$\ell ^{2}$ and $\alpha _{d}\left(
n_{1,+}-n_{1,-}\right) ^{2}=0.029$ meV$.$ The easy-axis anisotropy and
capacitive energy are comparable in size.

\section{QUANTUM HALL ISING FERROMAGNETS}

In this section, we extend the Ising behavior of bilayer graphene\cite{luo}
to chirally-stacked multilayer graphene and discuss the hysteretic behavior
associated with the valley-pseudospin flip transitions at integer filling $%
\nu _{N}=1.$

\subsection{Valley Ising quantum ferromagnetism and hysteretic behavior}

\begin{figure}[tbph]
\includegraphics[scale=1.0]{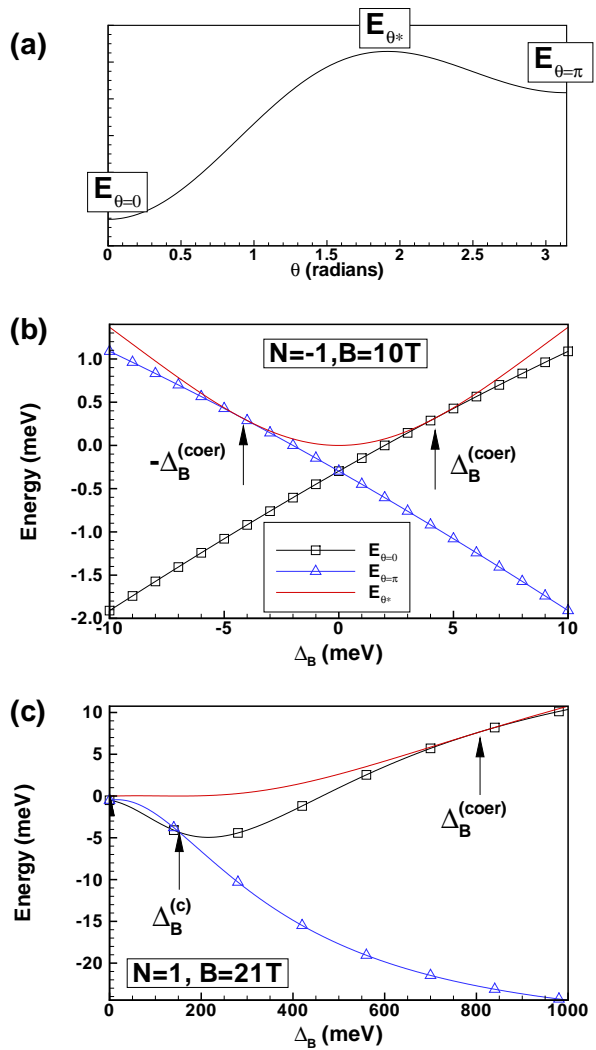}
\caption{(Color online) (a) General form of the energy functional $%
e_{HF}\left( \protect\theta \right) $ with the two minima at $\protect\theta %
=0,\protect\pi $ and the maximum at $\protect\theta ^{\ast }.$\ Energy per
electron at the extremal values of $e_{HF}\left( \protect\theta \right) $
for (b) $N=-1,B=10T;$ (c) $N=1,B=21$ T$.$ }
\label{Figure1}
\end{figure}

The ground state at $\nu _{N}=1$ is spin polarized and so, for a uniform
state, the energy functional of Eq. (\ref{totalener}) simplifies to (with $%
P_{z}\left( \mathbf{q}=0\right) =\frac{1}{2}\cos \theta $)%
\begin{equation}
e_{HF}\left( \theta \right) =\frac{E_{HF}\left( \theta \right) }{N_{e}}%
=C+\alpha \cos \left( \theta \right) +\beta \cos ^{2}\left( \theta \right) ,
\label{func}
\end{equation}%
where $C$ is a constant and $\alpha ,\beta ,$ which depend implicitly on the
bias $\Delta _{B},$ are given by 
\begin{equation}
\alpha =\frac{1}{2}\sum_{\xi }\left( \xi E_{\xi }^{\left( 0\right) }-\frac{1%
}{2}\xi X^{\left( \xi ,\xi \right) }-\xi \alpha _{d}n_{1,\xi }n_{2,\xi
}\right)  \label{alpha}
\end{equation}%
and%
\begin{eqnarray}
\beta &=&\frac{\alpha _{d}}{4}\left( n_{1,+}-n_{1,-}\right) ^{2} \\
&&-\frac{1}{8}\left( X^{\left( +,+\right) }+X^{\left( -,-\right)
}-2X^{\left( +,-\right) }\right)  \notag
\end{eqnarray}%
with all interactions evaluated at $\mathbf{q}=0.$ The form of the energy
functional $e_{HF}\left( \theta \right) $ (with $C$ neglected) is shown in
Fig. \ref{Figure1}(a) for a finite bias. There are two extrema at $\theta =0$
and $\theta =\pi $ and another one at $\cos \left( \theta ^{\ast }\right)
=-\alpha /2\beta $ with respective energies%
\begin{eqnarray}
E_{\theta =0} &=&\alpha +\beta , \\
E_{\theta =\pi } &=&-\alpha +\beta , \\
E_{\theta ^{\ast }} &=&-\alpha ^{2}/4\beta .
\end{eqnarray}%
The metastable minimum (chosen to be $\theta =\pi $ in the figure)
disappears when the curvature of $e_{HF}\left( \theta \right) $ changes sign
or, equivalently, when $E_{\theta =0}=E_{\theta ^{\ast }}$ or $E_{\theta
=\pi }=E_{\theta ^{\ast }}.$ We show in Fig. \ref{Figure1}(b) and (c) the
behavior of these three energies for $N=-1$ at $B=10$ T and $N=1$ at $B=21$
T.

When $\alpha =0$ and $\beta <0,$ or equivalently,%
\begin{equation}
X^{\left( +,+\right) }-X^{\left( +,-\right) }>\alpha _{d}\left(
n_{1,+}-n_{1,-}\right) ^{2},  \label{ising}
\end{equation}%
the two states $P_{z}\left( 0\right) =\pm \frac{1}{2}$ are degenerate and
the C2DEG has an Ising behavior$.$ This situation occurs at zero bias.

Numerically, the coefficient $\beta <0$ at all bias but $\alpha \neq 0$. As
Fig. \ref{Figure1} shows, $E_{\theta =0},E_{\theta =\pi }<E_{\theta ^{\ast
}} $ so that the Ising behavior persists to finite bias although the two
minima at $\theta =0,\pi $ then have unequal energies. One minimum is a
metastable state and the other a stable state. A hysteretic behavior is
expected since the system can be trapped into the metastable state if the
thermal energy is not sufficient to overcome the energy barrier between the
two minima. A similar behavior was discussed in semiconductor Ising QHFs\cite%
{jungwirthprb}, but the hysteretic behavior found in bilayer graphene for
valley pseudospins seems more complex.

For $N=-1,$ the coefficient $\alpha =0$ only at zero bias where there is a
transition from $P_{z}=\frac{1}{2}$ ($\Delta _{B}<0$) to $P_{z}=-\frac{1}{2}$
($\Delta _{B}>0$). In the ground state, all electrons are in the valley
which has the \textit{lowest} non-interacting energy $E_{\xi }^{\left(
0\right) }$. This is the situation depicted in Fig. \ref{Figure1}(b). For
positive bias, the metastable state at $\theta =0$ disappears at the bias $%
\Delta _{B}^{\left( \mathrm{coer}\right) }=4.2$ meV where $E_{\theta
=0}=E_{\theta ^{\ast }}$. In analogy with a magnetic system, $\Delta
_{B}^{\left( \mathrm{coer}\right) }$ can be considered as a coercive bias.
For negative bias, $\theta =\pi $ is the metastable state and a similar
scenario occurs. Hereafter, we consider positive bias only since the
transitions for negative bias are the opposite of those for positive bias.
All Landau levels $N<-1$ behave in the same fashion. The smallness of $%
\Delta _{B}^{\left( \mathrm{coer}\right) }$ for $N=-1$ reflects the fact
that $\alpha $ increases very rapidly with bias.

For $N=1,$ the ground state has $P_{z}=\frac{1}{2}$ at finite bias (the
electrons are in the valley state with the \textit{biggest} non-interacting
energy) until a critical bias $\Delta _{B}^{\left( c\right) }=152$ meV
indicated in Fig. \ref{Figure1}(c) where there is a jump from $P_{z}=+\frac{1%
}{2}$ to $P_{z}=-\frac{1}{2}$. Unlike $N=-1,$ the two minima in $%
e_{HF}\left( \theta \right) $ coexist up to a very large coercive bias $%
\Delta _{B}^{\left( \mathrm{coer}\right) }=808$ meV that is well outside the
limit of validity of our model that assumes no Landau level mixing. In order
to find the correct coercive bias for $N=1,$ it is necessary to include this
mixing.

We discuss the hysteretic loop for $N=\pm 1$ below (see Fig. \ref{Figure3}
(c)), after we obtain the gap for the bulk and confined electron-hole
excitations.

\subsection{Ising behavior in multilayer graphene}

The Ising behavior can be extended to chirally-stacked $m-$layer graphene ($%
m>1$), at least at zero bias. In the simplest tight-binding model where only 
$\gamma _{0}$ and $\gamma _{1}$ are kept and in the two-component model\cite%
{mccann}, the Hamiltonian in the basis $\left( A_{1},B_{m}\right) $ is 
\begin{equation}
H_{\xi }=a^{m-1}\alpha _{0}\xi ^{m}\left( 
\begin{array}{cc}
0 & \left( a^{\xi }\right) ^{m} \\ 
\left( a^{-\xi }\right) ^{m} & 0%
\end{array}%
\right) ,
\end{equation}%
where $a=\alpha _{0}/\gamma _{1}.$ The eigenvectors are of the form 
\begin{equation}
\frac{1}{\sqrt{2}}\left( 
\begin{array}{c}
h_{n}\left( \mathbf{r}\right) \\ 
c_{m}h_{n-m}\left( \mathbf{r}\right)%
\end{array}%
\right) ,
\end{equation}%
with $\left\vert c_{m}\right\vert ^{2}=1.$ (Landau level $\left\vert
N\right\vert =1$ corresponds to $n=m.$) This model can be captured by our
approach if only two-coefficients in the spinors of Eqs. (\ref{spinor1},\ref%
{spinor2}) are kept. This implies, from Eqs. (\ref{yy0}-\ref{yy5}), that $%
n_{1,\pm }=n_{2,\mp }=1/2.$ The Ising criteria in Eq. (\ref{ising}), becomes%
\begin{equation}
X^{\left( +,+\right) }-X^{\left( +,-\right) }>0
\end{equation}%
or 
\begin{equation}
\int_{0}^{\infty }dye^{\frac{-y^{2}}{2}}A^{2}\left( y\right) \left(
1-e^{-\left( m-1\right) yd/\ell }\right) >0,
\end{equation}%
with%
\begin{equation}
A\left( y\right) =L_{n}^{0}\left( \frac{y^{2}}{2}\right) -L_{n-m}^{0}\left( 
\frac{y^{2}}{2}\right) ,
\end{equation}%
where $d$ is the separation between two adjacent layers. This inequality is
satisfied for $m>1$ in chirally-stacked multilayer graphene.

\section{EXCITATION GAP FOR BULK ELECTRON-HOLE PAIRS}

We now study the charged excitations of the Ising QHFs at $\nu _{N}=1$ in
Landau levels $N=\pm 1.$ We start in this section with the Hartree-Fock bulk
quasiparticles (electron and hole) and consider in the next two sections the
Hartree-Fock electron and hole quasiparticles confined to the coherent
region of a valley-pseudospin domain wall and the spin and/or
valley-pseudospin skyrmions and antiskyrmions. In each case, we are mostly
interested in computing the energy to create a
quasiparticle-antiquasiparticle pair with infinite separation which is
simply the sum of the quasiparticle and antiquasiparticle energies. The
minimal such energy defines the transport gap of the Ising QHF which is
measurable in transport experiment.

In the uniform Ising phases, the equation of motion for the matrix of
Green's functions (Eq. (\ref{eqmotion})) gives four energy levels. When the
ground state is $\mu =1$ (i.e. $\mu _{\xi }=1$)$,$ the only nonzero order
parameter is $\left\langle \rho _{1,1}\left( 0\right) \right\rangle =1$
while when $\mu =3$ (i.e. $\mu _{\xi }=-1$) the only parameter is $%
\left\langle \rho _{3,3}\left( 0\right) \right\rangle =1.$ The first energy
level is fully occupied and amongst the three possible bulk electron-hole
excitations, two are always lower in energy: a spin flip and a
valley-pseudospin flip. For a ground state in valley $\xi ,$ the spin and
valley flips have the energy

\begin{eqnarray}
\Delta _{\mathrm{spin},\xi } &=&\Delta _{Z}+X_{N}^{\left( \xi ,\xi \right) },
\label{vallespin} \\
\Delta _{\mathrm{valley},\xi } &=&E_{-\xi }^{\left( 0\right) }-E_{\xi
}^{\left( 0\right) }  \label{valleppin} \\
&&+A_{N}^{\left( \xi ,-\xi \right) }-A_{N}^{\left( \xi ,\xi \right)
}+X_{N}^{\left( \xi ,\xi \right) }.  \notag
\end{eqnarray}

\begin{figure}[tbph]
\includegraphics[scale=1.0]{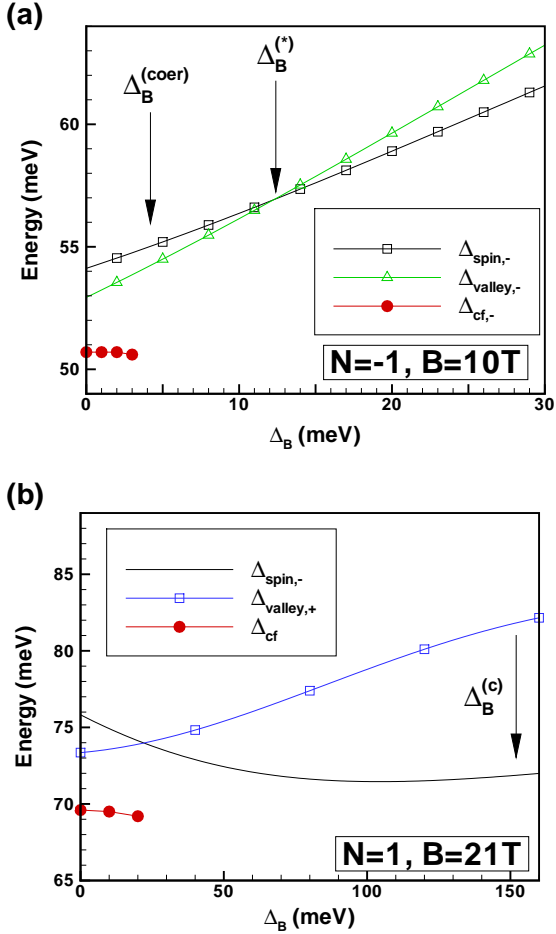}
\caption{(Color online) Energy of the bulk ($\Delta _{\mathrm{spin},\protect%
\xi }$ and $\Delta _{\mathrm{valley},\protect\xi }$) and confined ($\Delta _{%
\mathrm{cf},-}$) electron-hole excitations as a function of bias for: (a) $%
N=-1,B=10$ T and (b) $N=1,B=21$ T. }
\label{Figure2}
\end{figure}

These energies are plotted as a function of the bias in Fig. \ref{Figure2}%
(a) for $N=-1,B=10$ T and Fig. \ref{Figure2}(b) for $N=1,B=21$T. For $N=-1,$
the lowest-energy bulk excitation (i.e. the transport gap) is a
valley-pseudospin flip at small bias and it changes to a spin flip above $%
\Delta _{B}^{(\ast )}=12.2$ meV. There is a discontinuity in the slope of
the gap at $\Delta _{B}^{(\ast )}.$ For $N=1,$ the lowest-energy excitation
is also a valley-pseudospin flip until the critical bias $\Delta
_{B}^{\left( c\right) }=152$ meV. At this bias, a ground state transition
causes a discontinuity in the gap: it abruptly changes from a
valley-pseudospin flip to a spin flip. This is indicated by a the downward
arrow in Fig. \ref{Figure2}.

\section{EXCITATION GAP FOR ELECTRON-HOLE PAIRS CONFINED TO DOMAIN WALLS}

In the presence of disorder or at finite temperature, domain walls (DW) are
created in an Ising ferromagnet\cite{jungwirth}. These walls can have
different shapes, even closing on themselves (i.e. domain wall loops). In
this section, we find a straight domain wall solution to the microscopic
Hartree-Fock equation and compute the minimal energy required to excite an
electron-hole pair in its coherent region.

At filling factor $\nu _{N}=1,$ the ground state is spin polarized and all
electrons must be in valley $\xi =+1$ or $\xi =-1.$ For a state modulated in
only one direction, say along $x$, we can take $q_{y}=0$ in the definition
of the Fourier transform of the Green's function in Eq. (\ref{GX}) and so 
\begin{equation}
G_{\xi ,\xi ^{\prime }}\left( X,\tau \right) =-\left\langle Tc_{\xi
,X}\left( \tau \right) c_{\xi ^{\prime },X}^{\dagger }\left( 0\right)
\right\rangle .
\end{equation}%
The order parameters are now given by%
\begin{eqnarray}
\left\langle \rho _{\xi ,\xi ^{\prime }}\left( X\right) \right\rangle
&=&\left\langle c_{\xi ^{\prime },X}^{\dag }c_{\xi ,X}\right\rangle \\
&=&G_{\xi ^{\prime },\xi }\left( X,\tau =0^{-}\right) .  \notag
\end{eqnarray}%
The equation of motion (Eq. (\ref{eqmotion})) simplifies to%
\begin{eqnarray}
&&\left[ \hslash i\omega _{n}-\left( \widetilde{E}_{\xi }-\widetilde{\mu }%
\right) \right] G_{\xi ,\xi ^{\prime }}\left( X,\omega _{n}\right) =\hslash
\delta _{\xi ,\xi ^{\prime }}  \label{dwequation} \\
&&+H^{\xi }\left( X\right) G_{\xi ,\xi ^{\prime }}\left( X,\omega
_{n}\right) -\sum_{\xi ^{\prime \prime }}X^{\xi ,\xi ^{\prime \prime
}}\left( X\right) G_{\xi ^{\prime \prime },\xi ^{\prime }}\left( X,\omega
_{n}\right) ,  \notag
\end{eqnarray}%
with the renormalized energies%
\begin{equation}
\widetilde{E}_{\xi }=E_{\xi }^{\left( 0\right) }+\sum_{\zeta }A^{\left( \xi
,\zeta \right) }\left\langle \rho _{\zeta ,\zeta }\left( q_{x}=0\right)
\right\rangle ,
\end{equation}%
where 
\begin{equation}
\left\langle \widetilde{\rho }_{\xi ,\xi }\left( q_{x}\right) \right\rangle =%
\frac{1}{N_{\varphi }}\sum_{X}\left\langle \rho _{\xi ,\xi }\left( X\right)
\right\rangle e^{-iq_{x}X},
\end{equation}%
and the interactions 
\begin{eqnarray}
H^{\xi }\left( X\right) &=&\sum_{\xi ^{\prime \prime }}\overline{\sum_{q_{x}}%
}H^{\left( \xi ^{\prime \prime },\xi \right) }\left( q_{x}\right)
\label{hxx} \\
&&\times \left\langle \rho _{\xi ^{\prime \prime },\xi ^{\prime \prime
}}\left( q_{x}\right) \right\rangle e^{iq_{x}X},  \notag \\
X^{\xi ,\xi ^{\prime }}\left( X\right) &=&\sum_{q_{x}}X^{\left( \xi ^{\prime
},\xi \right) }\left( q_{x}\right) \left\langle \rho _{\xi ^{\prime },\xi
}\left( q_{x}\right) \right\rangle e^{iq_{x}X}.
\end{eqnarray}%
The Hartree-Fock energy per electron becomes 
\begin{eqnarray}
\frac{E_{DW}}{N_{e}} &=&\sum_{\xi }E_{\xi }^{\left( 0\right) }\left\langle 
\widetilde{\rho }_{\xi ,\xi }\left( 0\right) \right\rangle  \label{hx} \\
&&+\alpha _{d}\left( \left\langle \widetilde{\rho }_{1}\left( 0\right)
\right\rangle -\left\langle \widetilde{\rho }_{2}\left( 0\right)
\right\rangle \right) ^{2}  \notag \\
&&+\frac{1}{2N_{\varphi }}\sum_{\xi }\sum_{X}H^{\xi }\left( X\right)
\left\langle \rho _{\xi ,\xi }\left( X\right) \right\rangle  \notag \\
&&-\frac{1}{2N_{\varphi }}\sum_{\xi ,\zeta }\sum_{X}X^{\left( \xi ,\zeta
\right) }\left( X\right) \left\langle \rho _{\zeta ,\xi }\left( X\right)
\right\rangle .  \notag
\end{eqnarray}

The formal solution of the Green's function equation gives the band
structure (we leave implicit the dependence on $X$ of the variables here)%
\begin{equation}
E_{\pm }=\frac{1}{2}\left[ \overline{E}_{+}+\overline{E}_{-}\pm \Delta ^{2}%
\right] ,
\end{equation}%
where%
\begin{equation}
\overline{E}_{\pm }=\widetilde{E}_{\pm }+H^{\pm }-X^{\pm ,\pm }
\end{equation}%
and%
\begin{equation}
\Delta ^{2}=\left( \overline{E}_{+}-\overline{E}_{-}\right) ^{2}+\left\vert
X^{\left( +,-\right) }\right\vert ^{2}.
\end{equation}

The order parameters are obtained from the self-consistent equations

\begin{eqnarray}
\left\langle \rho _{+,+}\right\rangle &=&\frac{\left( E_{-}-\overline{E}%
_{-}\right) ^{2}}{\left( E_{-}-\overline{E}_{-}\right) ^{2}+\left\vert
X^{\left( +,-\right) }\right\vert ^{2}}, \\
\left\langle \rho _{-,-}\right\rangle &=&\frac{\left\vert X^{\left(
+,-\right) }\right\vert ^{2}}{\left( E_{-}-\overline{E}_{-}\right)
^{2}+\left\vert X^{\left( +,-\right) }\right\vert ^{2}}, \\
\left\langle \rho _{+,-}\right\rangle &=&\left\langle \rho
_{-,+}\right\rangle ^{\ast } \\
&=&\frac{-X^{\left( -,+\right) }\left( E_{-}-\overline{E}_{-}\right) }{%
\left( E_{-}-\overline{E}_{-}\right) ^{2}+\left\vert X^{\left( +,-\right)
}\right\vert ^{2}}.  \notag
\end{eqnarray}%
These equations are constrained by the condition $\left\langle \rho
_{+,+}\left( X\right) \right\rangle +\left\langle \rho _{-,-}\left( X\right)
\right\rangle =1$ so that there is no modulation of the \textit{total}
guiding-center density when the C2DEG is modulated in one direction only.
The electron-hole gap for the confined excitations is given by the minimal
separation between the two bands i.e. by 
\begin{equation}
\Delta _{\mathrm{cf}}=\mathrm{Min}\left[ E_{+}\left( X\right) -E_{-}\left(
X\right) \right] .
\end{equation}

We look for a DW solution by imposing the boudary conditions: $\left\langle
\rho _{+,+}\left( X=-\infty \right) \right\rangle =1$ and $\left\langle \rho
_{+,+}\left( X=\infty \right) \right\rangle =0.$ We start the iteration
process with the simple seed%
\begin{eqnarray}
\left\langle \rho _{\pm ,\pm }\left( X\right) \right\rangle &=&\frac{1}{2}%
\mp \frac{X}{L_{x}},  \label{s1} \\
\left\vert \left\langle \rho _{\pm ,\mp }\left( X\right) \right\rangle
\right\vert ^{2} &=&\frac{1}{4}-\frac{X^{2}}{L_{x}^{2}},  \label{s2}
\end{eqnarray}%
where $L_{x}=N_{\varphi }\Delta X$ is the length of the system in the
direction of the modulation and $N_{\varphi }$ the number of values of $X.$
We take both parameters as large as possible in the numerical calculation.
This seed produces a N\'{e}el wall. Multiplying the right-hand side of Eqs. (%
\ref{s1}) and (\ref{s2}) by $i$ produces a Bloch wall. Both walls have the
same energy. To represent the solution, we use a pseudospin representation
where%
\begin{eqnarray}
p_{x}+ip_{y} &=&2\left\langle \rho _{+,-}\right\rangle , \\
p_{z} &=&\left\langle \rho _{+,+}\right\rangle -\left\langle \rho
_{-,-}\right\rangle ,
\end{eqnarray}%
so that $\mathbf{p}\left( X\right) $ is a unit field in $X$-space.

\begin{figure}[tbph]
\includegraphics[scale=1.0]{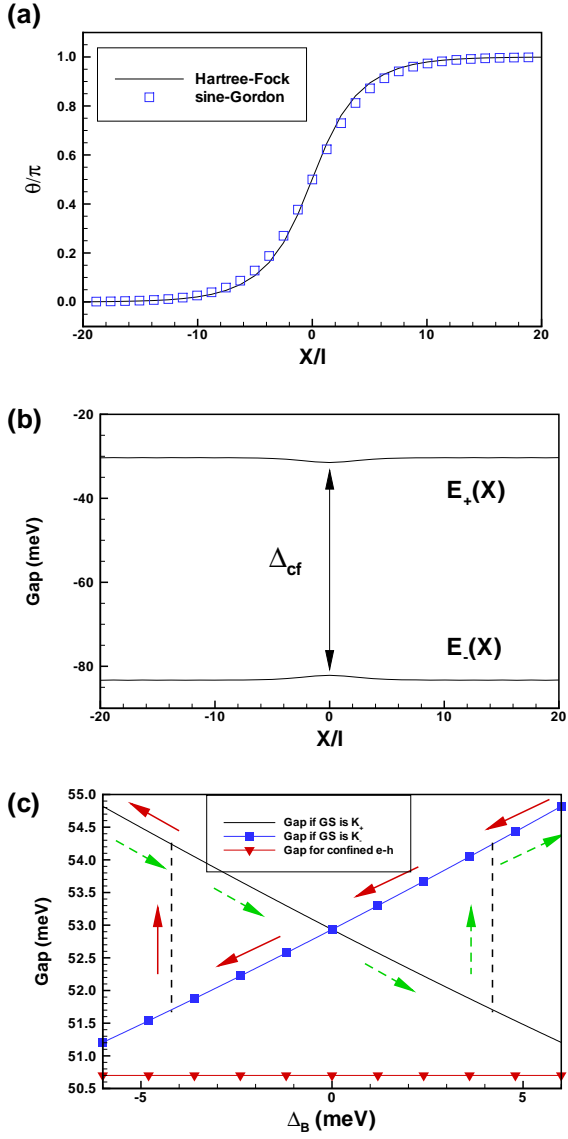}
\caption{(Color online) Domain wall for $\Delta _{B}=0,N=-1,B=10T$ and
hysteretic behavior. (a)\ Comparison of the sine-Gordon and Hartree-Fock
solutions. (b) Band structure, $E_{\pm }\left( X\right) $ in the domain wall
solution. The gap for a confined electron-hole excitation is indicated by $%
\Delta _{\mathrm{cf}}.$ (c) Hysteretic behavior of the transport gap around
zero bias. The plain (dashed) arrows indicate the progression of the gap for
a downward (upward) sweep in bias.}
\label{Figure3}
\end{figure}

Figure \ref{Figure3}(a)\ shows the profile of the angle $\theta \left(
X\right) $ between the pseudospin and the $z$ axis (i.e. $p_{z}\left(
X\right) =\cos \left[ \theta \left( X\right) \right] $) in the DW solution
at zero bias for $N=-1,$ and $B=10$ T$.$ The corresponding band structure is
shown in Fig. \ref{Figure3}(b) where the confined quasiparticle gap is
indicated by a double arrow. The intervalley coherence $\left\langle \rho
_{+,-}\left( X\right) \right\rangle $ is very strong near $X=0$ so that the
gap for the confined electron-hole pairs is smaller than the electron-hole
bulk gap by a small amount only. Bulk and confined gaps as a function of
bias are compared in Fig. \ref{Figure2}. Since the spin is not considered in
our DW solution, $\Delta _{\mathrm{cf}}$ should be compared with the
valley-pseudospin gap only in this figure. As the bias is increased
positively, the coherence region of the DW for $N=-1$ in Fig. \ref{Figure3}%
(a)\ moves to the left, indicating that the proportion of the stable $\xi
=-1 $ phase in the DW solution increases. At the same time, the metastable
minimum increases in energy. For $\Delta _{B}>3$ meV, only the $\xi =-1$
phase survives and there is no DW solution anymore. The DW solution
disappears before the coercive bias $\Delta _{B}^{\left( \mathrm{coer}%
\right) }=4.2$ meV is reached.

For $N=1,$ there is a DW solution up to $\Delta _{B}=20$ meV, a value much
lower than $\Delta _{B}^{(\mathrm{coer})}=808$ meV. Already at $\Delta
_{B}=20$ meV, the proportion of the stable phase $\xi =+1$ in the DW has
reached $100\%.$ No DW solution is found at larger bias until near the
critical bias $\Delta _{B}^{\left( c\right) }=152$ meV. At this bias, the
two Ising states have again the same energy (i.e. $\alpha =0$). Our
numerical analysis shows that this second DW solution exists in the range $%
\Delta _{B}\in \left[ 140,160\right] $ meV. Like the case $N=-1,$ the DW
solution is lost before the coercive bias is reached. The critical bias $%
\Delta _{B}^{\left( c\right) }$ is above the limit of validity of our model
but it is possible to modify either the magnetic field or the dielectric
constant to find a smaller $\Delta _{B}^{\left( c\right) }.$

Figure \ref{Figure3}(c) shows the hysteretic loop expected around zero bias
for $N=-1$. The dotted lines indicate the position of the coercive bias. If
the confined quasiparticles are not taken into account, then the arrows
indicate the progression of the gap for a downward or upward sweep in bias.
For a downward sweep, the gap decreases and then jumps suddenly, at $-\Delta
_{B}^{\left( c\right) },$ to its value in the phase where the electrons
occupy the valley $K_{+}.$ Just the opposite occurs for an upward sweep. If
skyrmions are considered, this behavior is not changed substantially since
the energy of a skyrmion-antiskyrmion pair is slightly different from that
of the bulk quasiparticles only in a small range of bias (see Sec. VII). If
confined quasiparticles are considered however (i.e. when domain walls are
present), then the gap jumps discontinuously at $\pm \Delta _{B}^{\left(
c\right) }$ and its value remains almost constant for $\left\vert \Delta
_{B}\right\vert <\Delta _{B}^{\left( c\right) }.$ The large value of the
coercive field for $N=1,$ if it survives Landau-level-mixing corrections,
should lead to a large hysteretic loop.

In the absence of interaction, the transport gap is zero at zero bias and
the QHE is lost if $\nu _{N}=1.$ The exchange interaction, however makes the
gap finite. In this case, the QHE is lost then when the gap is smaller than
the disorder broadening of the Landau levels.

It is interesting to compare the Hartree-Fock solution with the DW solution
that can be obtained by solving the Euler-Lagrange equation\cite{solyom}
derived from the energy functional of Eq. (\ref{evallee}). Neglecting the
capacitive term and at zero bias, the excitation energy is 
\begin{eqnarray}
\delta E_{\mathrm{valley}} &=&\frac{1}{2}L_{y}\int dx\left[ \rho _{z}\sin
^{2}\theta +\rho _{\bot }\cos ^{2}\theta \right] \left( \partial _{x}\theta
\right) ^{2} \\
&&+KL_{y}\int dx\cos ^{2}\theta .  \notag
\end{eqnarray}%
Considering a N\'{e}el wall and neglecting the small difference between the
stiffnesses $\rho _{z}$ and $\rho _{\bot }$ at zero bias, the Euler-Lagrange
equation simplifies to the sine-Gordon equation%
\begin{equation}
\frac{\partial ^{2}\varphi }{\partial x^{2}}=\frac{1}{\lambda ^{2}}\sin
\left( \varphi \right) ,
\end{equation}%
where $\varphi =2\theta $ and 
\begin{equation}
\lambda =\sqrt{\frac{\rho _{z}}{2\left\vert K\right\vert }}.
\end{equation}%
Imposing the boundary conditions $\theta \left( x=-\infty \right) =0$ and $%
\theta \left( x=\infty \right) =\pi ,$ a solution that minimizes the energy
is the sine-Gordon soliton\cite{rajaraman} 
\begin{equation}
\theta _{\mathrm{SG}}\left( x\right) =2\arctan \left( e^{\frac{x}{\lambda }%
}\right) ,  \label{gordon}
\end{equation}%
where $\lambda $ is the domain wall width (i.e. the coherence region). Note
that, in the sine-Gordon solution, $p_{z}(q_{x}=0)=0$ and the capacitive
term can effectively be neglected.

In Fig. \ref{Figure3}(a), the Hartree-Fock and sine-Gordon DW\ solutions for 
$N=-1,\Delta _{B}=0$ and $B=10$ T are compared. The sine-Gordon solution is
actually very close to the Hartree-Fock result. The domain wall width
extracted from Eq. (\ref{gordon}) is $\lambda =3.16\ell $. This is much
larger than the domain wall width found in AlAs quantum wells\cite{jungwirth}%
.

\section{EXCITATION GAP FOR SKYRMION-ANTISKYRMION PAIRS}

\subsection{Hamiltonian in the symmetric gauge}

To compute the skyrmion excitations, it is more convenient to use the
symmetric gauge $\mathbf{A}=\left( By/2,-Bx/2\right) $ with $\mathbf{B}=-B%
\widehat{z}$ instead of the Landau gauge. The Landau gauge wave functions $%
h_{n,X}\left( \mathbf{r}\right) $ in the spinors of Eqs. (\ref{spinor1})-(%
\ref{spinor2}) are then be replaced by the functions $\Phi _{n,m}\left( 
\mathbf{r}\right) $ where the quantum number $m=0,1,2,...$ is related to the
angular momentum $l_{z}$ by $l_{z}=\left( m-n\right) \hslash .$ The
functions $\Phi _{n,m}\left( \mathbf{r}\right) $ are given by\cite{yoshioka}
(by definition, $\Phi _{n,m}\left( \mathbf{r}\right) =0$ if $n<0$ or $m<0$)%
\begin{eqnarray}
\Phi _{n,m}\left( \mathbf{r}\right) &=&B_{n,m}e^{i\left( m-n\right) \varphi
}\left( \frac{r}{\ell }\right) ^{\left\vert m-n\right\vert }e^{-\frac{r^{2}}{%
4\ell ^{2}}} \\
&&\times L_{\frac{n+m}{2}-\frac{\left\vert n-m\right\vert }{2}}^{\left\vert
m-n\right\vert }\left( \frac{r^{2}}{2\ell ^{2}}\right) ,  \notag
\end{eqnarray}%
where $\varphi $ is the angle between the vector $\mathbf{r}$ and the $x$
axis, $L_{n}^{m}\left( x\right) $ is a generalized Laguerre polynomial and
the normalization constant is given by%
\begin{equation}
B_{n,m}=\frac{C_{n,m}\left( -i\right) ^{n}}{\sqrt{2^{\left\vert
m-n\right\vert }2\pi \ell ^{2}}}\sqrt{\frac{\left( \frac{n+m}{2}-\frac{%
\left\vert n-m\right\vert }{2}\right) !}{\left( \frac{n+m}{2}+\frac{%
\left\vert n-m\right\vert }{2}\right) !}}
\end{equation}%
with\cite{note3} $C_{n,m}=1$ if $m\geq n$ and $C_{n,m}=\left( -1\right)
^{m-n}$ if $m<n$.

In the Hartree-Fock approximation, the Hamiltonian in the symmetric gauge is
given by 
\begin{eqnarray}
H_{HF} &=&\sum_{\mu ,m}\left( E_{\mu _{\xi }}^{\left( 0\right) }-\mu
_{\sigma }\Delta _{Z}\right) c_{\mu ,m}^{\dag }c_{\mu ,m}  \label{HHF} \\
&&-\sum_{\mu }\sum_{m_{1},m_{2}}U_{m_{1},m_{1},m_{2},m_{2}}^{\gamma _{\xi
},\gamma _{\xi },\mu _{\xi },\mu _{\xi }}c_{\mu ,m_{2}}^{\dag }c_{\mu ,m_{2}}
\notag \\
&&+\sum_{\mu ,\nu }\sum_{m_{1},\ldots
,m_{4}}U_{m_{1},m_{2},m_{3},m_{4}}^{\mu _{\xi },\mu _{\xi },\nu _{\xi },\nu
_{\xi }}\left( \left\langle c_{\mu ,m_{1}}^{\dag }c_{\mu
,m_{2}}\right\rangle \right.  \notag \\
&&\left. \times c_{\nu ,m_{3}}^{\dag }c_{\nu ,m_{4}}-\left\langle c_{\mu
,m_{1}}^{\dag }c_{\nu ,m_{4}}\right\rangle c_{\nu ,m_{3}}^{\dag }c_{\mu
,m_{2}}\right) ,  \notag
\end{eqnarray}%
where the matrix elements of the Coulomb interaction are defined by%
\begin{eqnarray}
U_{m_{1},m_{2},m_{3},m_{4}}^{\mu _{\xi },\mu _{\xi },\nu _{\xi },\nu _{\xi
}} &=&\sum_{i,j=1}^{4}\left\vert b_{\mu _{\xi },i}\right\vert ^{2}\left\vert
b_{\nu _{\xi },j}\right\vert ^{2} \\
&&\times V\left( i,j\right) _{m_{1},m_{2},m_{3},m_{4}}^{\left( \alpha _{\mu
_{\xi }}\right) _{i},\left( \alpha _{\mu _{\xi }}\right) _{i},\left( \alpha
_{\nu _{\xi }}\right) _{j},\left( \alpha _{\nu _{\xi }}\right) _{j}},  \notag
\end{eqnarray}%
with%
\begin{eqnarray}
\alpha _{-1} &=&\left( n-1,n,n-2,n-1\right) , \\
\alpha _{+1} &=&\left( n-1,n-2,n,n-1\right) ,
\end{eqnarray}%
and 
\begin{eqnarray}
&&V\left( i,j\right) _{m_{1},m_{2},m_{3},m_{4}}^{n,n,n^{\prime },n^{\prime }}
\\
&=&\left( \frac{e^{2}}{\kappa \ell }\right) \delta _{m_{1}-m_{2}+m_{4}-m_{3}}
\notag \\
&&\times \sqrt{\frac{\min \left( m_{1},m_{2}\right) !\min \left(
m_{3},m_{4}\right) !}{\max \left( m_{1},m_{2}\right) !\max \left(
m_{3},m_{4}\right) !}}  \notag \\
&&\times \int_{0}^{\infty }dke^{-k^{2}}e^{-\Delta \left( i,j\right) kd/\ell
}\left( \frac{k^{2}}{2}\right) ^{\left\vert m_{1}-m_{2}\right\vert
}L_{n}^{0}\left( \frac{k^{2}}{2}\right)  \notag \\
&&\times L_{n^{\prime }}^{0}\left( \frac{k^{2}}{2}\right) L_{\min
(m_{1},m_{2})}^{\left\vert m_{1}-m_{2}\right\vert }\left( \frac{k^{2}}{2}%
\right) L_{\min (m_{3},m_{4})}^{\left\vert m_{3}-m_{4}\right\vert }\left( 
\frac{k^{2}}{2}\right) ,  \notag
\end{eqnarray}%
where $\Delta \left( i,j\right) =0$ or $1$ when $\left\vert i-j\right\vert
\leq 1$ or $\left\vert i-j\right\vert \geq 2.$ The Hamiltonian $H_{HF}$
includes an interaction with a positive background where the positive
charges are assumed to occupy the same state, $\gamma ,$ as the electrons in
the ground state.

\subsection{Spin and valley-pseudospin skyrmions}

At $\nu _{N}=1,$ the ground state has $\mu =1$ or $\mu =3$ and can be
written as 
\begin{equation}
\left\vert \mathrm{GS}\right\rangle =\prod\limits_{m=0}c_{\mu ,m}^{\dag
}\left\vert 0\right\rangle .
\end{equation}%
If, for example, $\mu =1,$ then a general antiskyrmion (one electron removed
from the ground state) that allows for the possibility of an intertwined
spin and valley-pseudospin textures can be written as 
\begin{equation}
\left\vert \mathrm{aSk}\right\rangle =\prod\limits_{m=0}\xi _{m}^{\dag
}\left\vert 0\right\rangle ,  \label{a1}
\end{equation}%
where the operator%
\begin{eqnarray}
\xi _{m}^{\dag } &=&u_{1,m}c_{1,m+1}^{\dag }+u_{2,m}c_{2,m}^{\dag }
\label{l1} \\
&&+u_{3,m}c_{3,m}^{\dag }+u_{4,m}c_{4,m}^{\dag }.  \notag
\end{eqnarray}%
For a skyrmion (one electron added to the ground state), three types of
excitations are possible\cite{ezawa} depending on the level the extra
electron is added to i.e.: 
\begin{equation}
\left\vert \mathrm{Sk}\right\rangle _{k}=\prod\limits_{m=0}\xi _{k,m}^{\dag
}c_{k,0}^{\dag }\left\vert 0\right\rangle ,  \label{a2}
\end{equation}%
with $k=2,3,4$ and%
\begin{eqnarray}
\xi _{2,m}^{\dag } &=&u_{1,m}c_{1,m}^{\dag }+u_{2,m}c_{2,m+1}^{\dag }
\label{l2} \\
&&+u_{3,m}c_{3,m}^{\dag }+u_{4,m}c_{4,m}^{\dag },  \notag
\end{eqnarray}%
\begin{eqnarray}
\xi _{3,m}^{\dag } &=&u_{1,m}c_{1,m}^{\dag }+u_{2,m}c_{2,m}^{\dag }
\label{l3} \\
&&+u_{3,m}c_{3,m+1}^{\dag }+u_{4,m}c_{4,m}^{\dag },  \notag
\end{eqnarray}%
\begin{eqnarray}
\xi _{4,m}^{\dag } &=&u_{1,m}c_{1,m}^{\dag }+u_{2,m}c_{2,m}^{\dag }
\label{l4} \\
&&+u_{3,m}c_{3,m}^{\dag }+u_{4,m}c_{4,m+1}^{\dag }.  \notag
\end{eqnarray}%
The normalization condition imposes $\sum_{\mu }\left\vert u_{\mu
,m}\right\vert ^{2}=1$ for each $m.$ A similar scheme is applied if the
ground state has $\mu =3.$

When the coefficients $u_{\mu ,m}=0$ for $\mu =2,3,4,$ the antiskyrmion or
skyrmion excitations are reduced to the bulk hole $\left\vert \mathrm{h}%
\right\rangle =c_{1,0}\left\vert \mathrm{GS}\right\rangle $ or bulk electron
excitations $\left\vert \mathrm{e}\right\rangle _{k}=c_{k,0}^{\dag
}\left\vert \mathrm{GS}\right\rangle $ respectively. If $u_{2,m}=u_{4,m}=0,$
the excitation is a pure valley-pseudospin skyrmion, $\left\vert \mathrm{VSk}%
\right\rangle ,$ or valley-pseudospin antiskyrmion, $\left\vert \mathrm{VaSk}%
\right\rangle ,$ with spin $\sigma =+1.$ If $u_{3,m}=u_{4,m}=0$ (or $%
u_{1,m}=u_{2,m}=0$), the excitation is a pure spin-skyrmion, $\left\vert 
\mathrm{SSk}\right\rangle ,$ or spin-antiskyrmion, $\left\vert \mathrm{SaSk}%
\right\rangle ,$ in valley $K_{+}$\ (or $K_{-}$).

To compute the coefficients $u_{p,m}$ for the antiskyrmion excitation, we
define the $4\times 4$ matrix of Green's functions $G_{m}\left( \tau \right)
=-\left\langle Tv_{m}\left( 0\right) v_{m}^{\dag }\left( \tau \right)
\right\rangle $ where $v_{m}=\left( 
\begin{array}{cccc}
c_{1,m+1} & c_{2,m} & c_{3,m} & c_{4,m}%
\end{array}%
\right) ^{\dag }$ and write down its Hartree-Fock equation of motion using
the Hamiltonian of Eq. (\ref{HHF}). The procedure is described in detail in
Ref. \onlinecite{luo2} and is equivalent to the canonical transformation
method used to compute spin-skyrmion in Ref. \onlinecite{fertig}. The
resulting self-consistent equation for the matrix Green's function is solved
numerically in an iterative way until a converging solution is found. We
proceed in a similar way for the three skyrmion excitations. The occupation
of each quantum state $m$ is given by $\left\vert u_{\mu ,m}\right\vert ^{2}$
and the coherence factors by $u_{\mu ,m}u_{\nu ,m}^{\ast }$ ($\mu \neq \nu $%
). These coherences are responsible for the spin and valley-pseudospin
textures. They are obtained from the off-diagonal elements of the matrix $%
G_{m}\left( \tau =0^{-}\right) $. In the numerical calculation, the maximum
value of $m$ is set to $m_{\max }=500$ so that skyrmions up to $R\approx 
\sqrt{2m_{\max }}\ell $ in size can be obtained.

For a pure spin skyrmion, the number of flipped spins per skyrmion when the
ground state has electrons with spin $\sigma =+1$ in valley $\xi =1$ or $-1$
is given by%
\begin{equation}
K_{\mathrm{spin}}=\sum_{m}\left\vert u_{2\text{ }\mathrm{or}\text{ }%
4,m}\right\vert ^{2},
\end{equation}%
while the number of flipped valley pseudospins per skyrmion is 
\begin{equation}
K_{\mathrm{pspin}}=\sum_{m}\left\vert u_{3\text{ }\mathrm{or}\text{ }%
1,m}\right\vert ^{2}.
\end{equation}

\subsection{Numerical results for skyrmions}

\begin{figure}[tbph]
\includegraphics[scale=1.0]{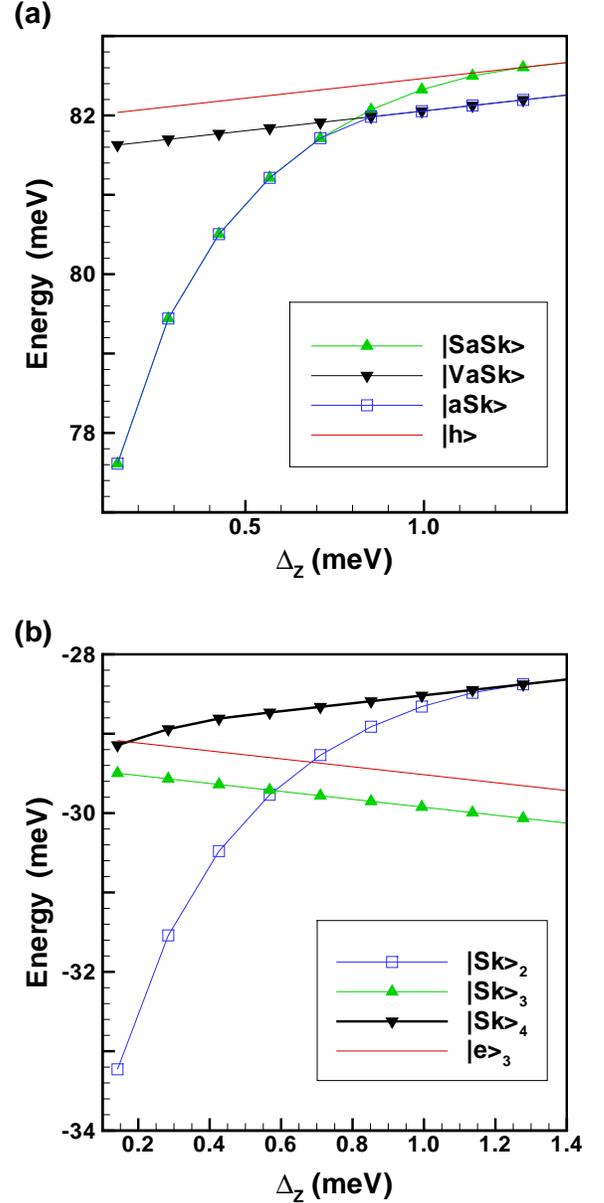}
\caption{(Color online) Excitation energy at zero bias of the (a)
antiskyrmion and bulk quasi-hole; (b) skyrmion and bulk quasi-electron as a
function of the Zeeman coupling for Landau level $N=-1$ and magnetic field $%
B=10$ T. }
\label{Figure4}
\end{figure}

\begin{figure}[tbph]
\includegraphics[scale=1.0]{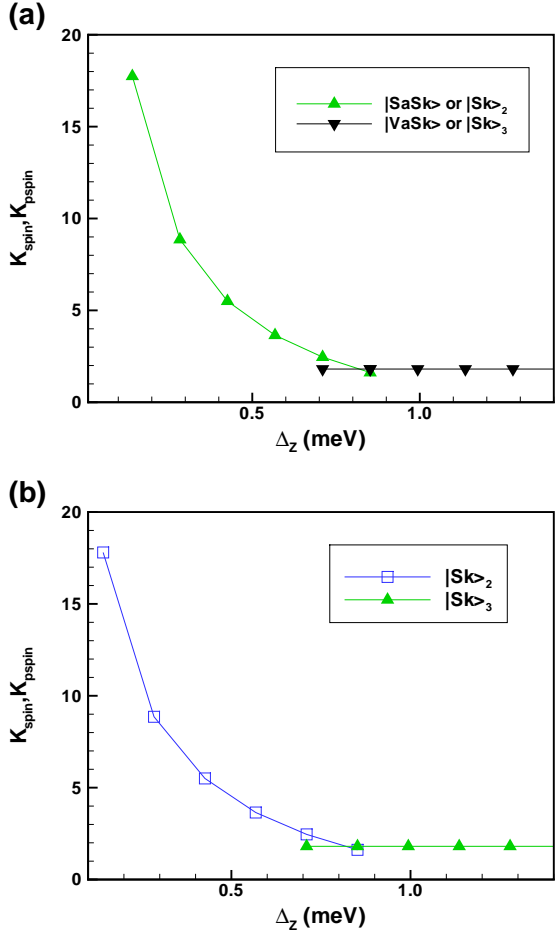}
\caption{(Color online) Number of spin flips ($K_{\mathrm{spin}}$) or valley
pseudospin flips ($K_{\mathrm{pspin}}$) as a function of the Zeeman coupling
for the lowest-energy excitations in Fig. 4. }
\label{Figure5}
\end{figure}

We choose $\mu =1$ for the ground state at zero bias. Figure \ref{Figure4}%
(a) shows the excitation energies of the antiskyrmion and bulk hole as a
function of the Zeeman coupling for Landau level $N=-1$. Note that, with a
maximum number of angular momenta $m_{\max }=500$, we are limited to the
range $\Delta _{Z}\geq 0.14$ meV\textbf{\ }otherwise the skyrmion is too big.%
\textbf{\ }Figure \ref{Figure4} shows that the antiskyrmion has lower energy
than the bulk hole in the small Zeeman range $\Delta _{Z}\in \left[ 0,1.28%
\right] $ meV. There is no intertwined spin-valley pseudospin texture in
this range. Indeed, for $\Delta _{Z}\lesssim 0.85$ meV, the excitation is a
pure spin antiskyrmion, $\left\vert \text{\textrm{SaSk}}\right\rangle ,$
while above this value it is a pure valley-pseudospin antiskyrmion,$%
\left\vert \text{\textrm{VaSk}}\right\rangle $. The corresponding number of
spin flips $K_{\mathrm{spin}}$ or valley-pseudospin flips $K_{\mathrm{pspin}%
} $ is shown in Fig. \ref{Figure5}(a) as a function of the Zeeman coupling.
Naturally, $K_{\mathrm{pspin}}$ is independent of the Zeeman coupling while $%
K_{\mathrm{spin}}$ increases with decreasing $\Delta _{Z}.$

Figure \ref{Figure4}(b) shows the excitation energy for the skyrmion and
bulk electron states for $N=-1$ and zero bias. The state $\left\vert \mathrm{%
e}\right\rangle _{3}$ has the lowest energy amongst the bulk electron states
since the two other bulk states $\left\vert \mathrm{e}\right\rangle _{2}$
and $\left\vert \mathrm{e}\right\rangle _{4}$ (not shown in the figure) are
degenerate and involve a spin flip. The lowest-energy skyrmion is $%
\left\vert \mathrm{Sk}\right\rangle _{2}$ for $\Delta _{Z}<0.59$ meV and $%
\left\vert \mathrm{Sk}\right\rangle _{3}$ above this value. The transition
between these two skyrmion types is discontinuous. Again, no intertwined
texture is found. The excitation $\left\vert \mathrm{Sk}\right\rangle _{2}$
and $\left\vert \mathrm{Sk}\right\rangle _{3}$ are respectively a spin and a
valley-pseudospin skyrmion. Apart from a global Zeeman shift of their energy
due to added or removed electron, the energy of the valley-pseudospin
skyrmion and antiskyrmion is independent of the Zeeman coupling since these
excitations do contain spin flip. Moreover, the coefficients of the
eigenspinors in Eqs. (\ref{spinor1}-\ref{spinor2}) do not depend on the
Zeeman coupling either. The number of spin flips for $\left\vert \mathrm{Sk}%
\right\rangle _{2}\left( \left\vert \mathrm{Sk}\right\rangle _{3}\right) $
is identical to that for $\left\vert \text{\textrm{SaSk}}\right\rangle
\left( \left\vert \text{\textrm{VaSk}}\right\rangle \right) $.

The sum of the skyrmion and antiskyrmion energies gives the energy needed to
create a skyrmion-antiskyrmion pair with an infinite separation. At zero
Zeeman coupling, the spin skyrmion-antiskyrmion pair has lower energy than
the bulk electron-hole pair in the range $\Delta _{Z}\in \left[ 0,1.28\right]
$ meV. In graphene however, $\Delta _{Z}=1.16$ meV if the magnetic field is
not tilted. At this value, the relevant excitations for the transport gap
(i.e. the pair excitations with the lowest energy) are the valley-pseudospin
skyrmion-antiskyrmion pairs.

\begin{figure}[tbph]
\includegraphics[scale=1.0]{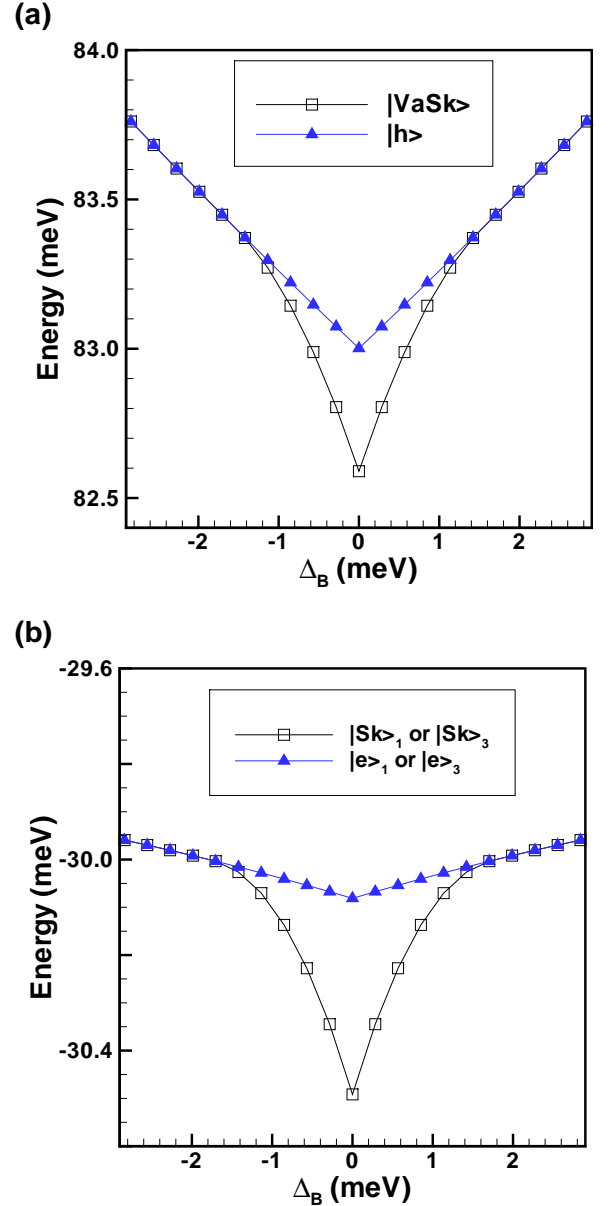}
\caption{(Color online) Excitation energy of (a) valley antiskyrmion and
quasi-hole; (b) valley skyrmion and quasi-electron for bias $\Delta
_{B}=0.03e^{2}/\protect\kappa \ell ,$ $B=10$ T and $N=-1.$ The excitations
are $\left\vert e\right\rangle _{1}$ and $\left\vert Sk\right\rangle _{1}$
for positive bias and $\left\vert e\right\rangle _{3}$ and $\left\vert
Sk\right\rangle _{3}$ for negative bias.}
\label{Figure6}
\end{figure}

\begin{figure}[tbph]
\includegraphics[scale=1.0]{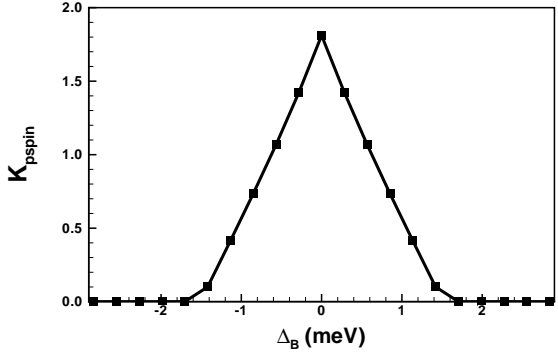}
\caption{Number of flipped valley pseudospins per skyrmion or antiskyrmion
as a function of bias in Landau level $N=-1$ and magnetic field $B=10$ T.}
\label{Figure7}
\end{figure}

To study the valley-pseudospin skyrmions as a function of bias, we take $%
\Delta _{Z}=2.1$ meV, i.e. a value large enough to kill the spin texture.
The ground state for $N\leq -1$ has $\mu =3$ for $\Delta _{B}>0$ and $\mu =1$
for $\Delta _{B}<0$. We find a pure valley-pseudospin skyrmion in this case: 
$\left\vert \mathrm{Sk}\right\rangle _{1}$ for $\Delta _{B}>0$ and $%
\left\vert \mathrm{Sk}\right\rangle _{3}$ for $\Delta _{B}<0.$ The
excitation energy of valley-pseudospin antiskyrmion and skyrmion is plotted
in Fig. \ref{Figure6} (a) and (b) together with the energy of the
corresponding bulk electron and hole. The valley-pseudospin texture
gradually disappears as $\left\vert \Delta _{B}\right\vert $ increases. The
range of bias where the topological excitations win over the bulk ones is
small: $\left\vert \Delta _{B}\right\vert \leq 1.77$ meV and $\left\vert
\Delta _{B}\right\vert \leq 1.42$ meV respectively. The number of
valley-pseudospin flips, $K_{\mathrm{pspin}}$ is shown in Fig. \ref{Figure7}%
. It is maximal at zero bias and decreases rapidly with increasing bias. At
zero bias, $K_{\mathrm{pspin}}\approx 2,$ and so these skyrmions are very
small in space unlike the spin skyrmions at zero Zeeman coupling.

The induced density profiles $\delta n\left( \mathbf{r}\right) =n\left( 
\mathbf{r}\right) -n_{GS}\left( \mathbf{r}\right) $ of the spin and
valley-pseudospin skyrmions or antiskyrmions are similar to that of spin
skyrmions in a semiconductor 2DEG\cite{ezawa}. The density $\delta n\left( 
\mathbf{r}\right) $ is maximal and positive at the origin for a skyrmion (or
minimal and negative for an antiskyrmion). It is isotropic and
decreases(increases) to zero at infinity for skyrmion(antiskyrmion). The
integrated density $\int d\mathbf{r}\delta n\left( \mathbf{r}\right) $
contains one more(less) electron for a skyrmion(antiskyrmion). A plot of the
valley-pseudospin texture in the projected representation, which is defined
in Appendix A, gives the pattern represented in Fig. \ref{Figure8}. The
pairing with $\Delta m=\pm 1$ in Eqs. (\ref{l1})-(\ref{l4}) leads to a spin
or valley-pseudospin vortex around the origin where the pseudospins rotate
by $\pm 2\pi $.

\begin{figure}[tbph]
\includegraphics[scale=1.0]{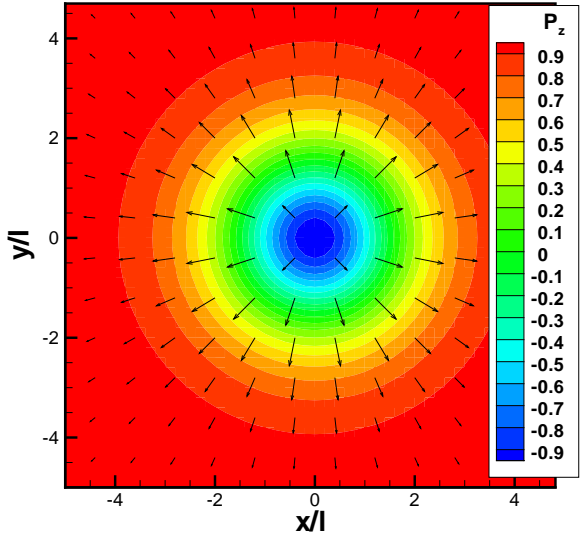}
\caption{(Color online) Valley-pseudospin texture of a valley-pseudospin
skyrmion in Landau level $N=-1$ and magnetic field $B=10$ T in the projected
representation.}
\label{Figure8}
\end{figure}

\begin{figure}[tbph]
\includegraphics[scale=1.0]{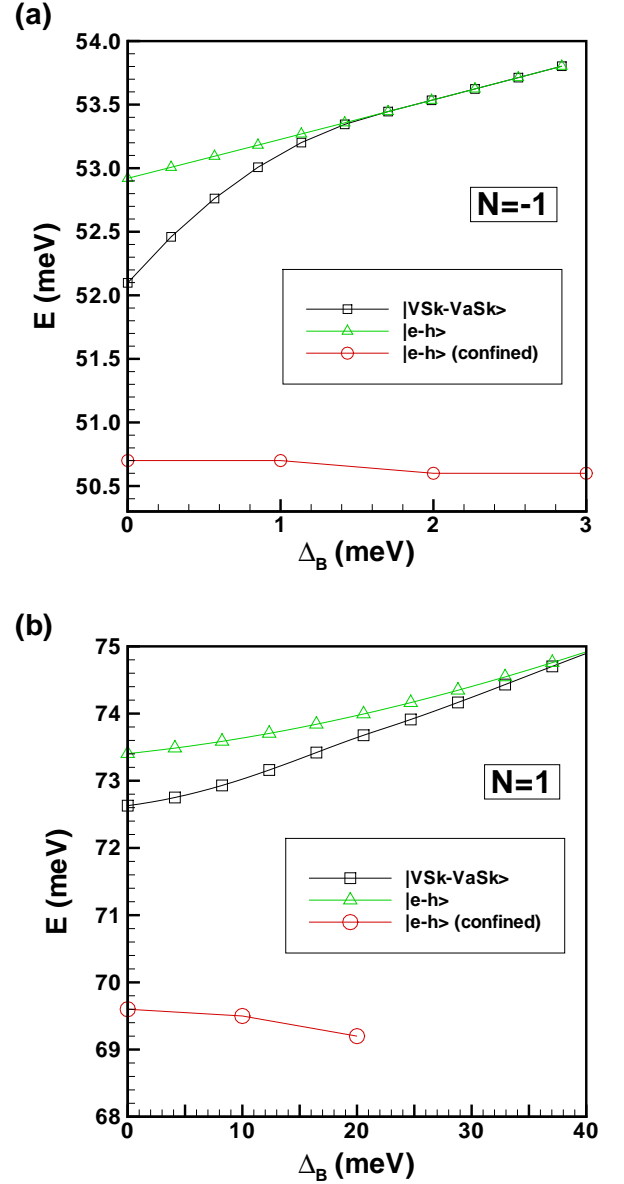}
\caption{(Color online) Energy with bias of bulk and confined electron-hole
pairs and of valley-pseudospin skyrmion-antiskyrmion pairs. (a) $N=-1,B=10$
T ; (b) $N=1,B=21$ T.}
\label{Figure9}
\end{figure}

Figure \ref{Figure9} shows the energy of the different pair excitations in
Landau level $N=-1$ with $B=10$ T and in level $N=1$ with $B=21$ T with the
Zeeman coupling $\Delta _{Z}=g\mu _{B}B$ in both cases. Spin
skyrmion-antiskyrmion pairs are not relevant at these Zeeman couplings and
so only the bulk and confined electron-hole pairs and the valley-pseudospin
skyrmion-antiskyrmion pairs are considered. From this figure, it is clear
that, if domain walls are present, then the confined pairs are the relevant
excitations at small bias. Otherwise, skyrmion-antiskyrmion pairs win but
their energy is only slightly lower than that of the bulk quasiparticles.

We were not able to obtain spin or valley-pseudospin skyrmions in Landau
levels $\left\vert N\right\vert >1$. For the spin skyrmions, it was found%
\cite{luo2} that in graphene (monolayer), the range of Zeeman coupling where
skyrmions are the lowest-energy excitations decreases very rapidly with
increasing $\left\vert N\right\vert .$ We assume that the same behavior
applies to bilayer graphene and that, if skyrmions exist (according to the
nonlinear $\sigma $ model discussed in the next section, they do for some
higher values of $\left\vert N\right\vert $), then they do so in a Zeeman
range not accessible to our numerical approach. For valley-pseudospin
skyrmions, the number of pseudospin flips shown in Fig. \ref{Figure8}
indicates that they are very small. Increasing the Landau level index will
make them even smaller, indistinguishable from the bulk charged excitations.

\subsection{Spin skyrmion in the gradient approximation}

It may seem surprising that the excitation energy of the spin skyrmion is
lower than that of the valley-pseudospin skyrmion at small Zeeman coupling
and zero bias (see Fig. \ref{Figure4}) since, in the non-interacting
picture, there is no cost for a valley pseudospin flip and the
exchange-energy cost seems similar in the two excitations. However, as shown
in Sec. III, the energy functional in the gradient approximation is very
different for spin and valley-pseudospin textures. For valley pseudospins,
the exchange interaction is anisotropic and there is an easy-axis anisotropy
term and an easy-plane term (the capacitive energy).

In the limit $\Delta _{Z}\rightarrow 0,$ skyrmions have a slowly-varying
spin texture and their excitation energy can be computed from the nonlinear $%
\sigma $ model (NL$\sigma $M) energy functional in Eq. (\ref{efunspin}). The
energy needed to create a spin skyrmion-antiskyrmion pair with infinite
separation is\cite{sondhi,moon}%
\begin{equation}
E_{SSk-SaSk}^{NL\sigma M}=8\pi \rho _{S},  \label{eskask}
\end{equation}%
where the spin stiffness is defined in Eq. (\ref{stifspin}). This value
provides a lower bound for the Hartree-Fock calculations shown in Fig. \ref%
{Figure4} since the limit $\Delta _{Z}\rightarrow 0$ cannot be reached
numerically. According to Eq. (\ref{eskask}), a spin skyrmion-antiskyrmion
pair at zero Zeeman and zero bias has an energy lower than a bulk
electron-hole pair for $N\in \left[ -3,-1\right] $ and $N\in \left[ 1,4%
\right] $ at $B=10$ T. Table 1 lists the value of these two gap energies for
different Landau levels.

\begin{center}
\begin{table}[tbp] \centering%
\begin{tabular}{|l|l|l|}
\hline
$N$ & $E_{\left( e-h\right) ,spin}^{Bulk}$ (meV) & $E_{SSk-SaSk}^{NL\sigma M}
$ (meV) \\ \hline
$\pm 1$ & $%
\begin{array}{c}
51.1 \\ 
52.9%
\end{array}%
$ & $%
\begin{array}{c}
24.2 \\ 
24.5%
\end{array}%
$ \\ \hline
$\pm 2$ & $%
\begin{array}{c}
43.2 \\ 
43.9%
\end{array}%
$ & $%
\begin{array}{c}
28.8 \\ 
31.4%
\end{array}%
$ \\ \hline
$\pm 3$ & $%
\begin{array}{c}
38.9 \\ 
39.3%
\end{array}%
$ & $%
\begin{array}{c}
31.5 \\ 
35.0%
\end{array}%
$ \\ \hline
$\pm 4$ & $%
\begin{array}{c}
36.0 \\ 
36.2%
\end{array}%
$ & $%
\begin{array}{c}
33.5 \\ 
37.4%
\end{array}%
$ \\ \hline
$\pm 5$ & $%
\begin{array}{c}
33.8 \\ 
34.0%
\end{array}%
$ & $%
\begin{array}{c}
35.3 \\ 
39.4%
\end{array}%
$ \\ \hline
\end{tabular}%
\caption{Energy of a bulk electron-hole pair and a spin skyrmion-antiskyrmion pair in the NL$\sigma$M calculated as a function
the Landau level index $N$ at zero bias and zero Zeeman coupling and at $B=10$ T. The results are the same in the two 
valleys. In each entry, the upper(lower) value is for the positive(negative) values of $N$.}%
\label{Table1}%
\end{table}%
\end{center}

\subsection{Valley skyrmion in the gradient approximation}

If the easy-axis anisotropy and capacitive terms could be neglected in the
anisotropic nonlinear $\sigma $ model (ANL$\sigma $M)\ energy functional of
Eq. (\ref{evallee}), then the valley-pseudospin skyrmion-antiskyrmion
excitation (or bimeron-antibimeron) energy would be given by \cite{brey}%
\begin{equation}
E_{VSk-VaSk}^{ANL\sigma M}=\frac{16\pi }{3}\rho _{\bot }+\frac{8\pi }{3}\rho
_{z}  \label{bimeron}
\end{equation}%
and the number of valley-pseudospin flips would diverge at zero bias. As our
microscopic calculations shows, this does not happen. Valley-pseudospin
skyrmions have a finite small size at zero bias and so Eq. (\ref{bimeron})
cannot be used to compute their energies. It is necessary to include the
capacitive and easy-axis term as well as the Hartree energy due to the
density modulation induced by the skyrmions. All these terms are included in
our microscopic Hartree-Fock calculation of the skyrmion energy.

Inter-Landau-level spin-skyrmions have been studied before in semiconductor
quantum wells\cite{lilliehook} but it was shown that are never the
lowest-lying charged excitations. By contrast, we find here, in bilayer
graphene, that skyrmions where different orbitals (but of the same Landau
level) are mixed can be the lowest-lying charged excitations.

\section{CONCLUSION}

We have computed the energy of different types of pair excitations in the
Ising quantum Hall ferromagnetic states of bilayer graphene in Landau levels 
$\left\vert N\right\vert >0$. These excitations include bulk electron-hole
pairs, electron-hole pairs confined to the coherent region of a domain wall,
and skyrmions-antiskyrmion pairs. For a disorder-free systems where the
Coulomb interaction is treated in the Hartree-Fock approximation,
valley-pseudospin skyrmion-antiskyrmion pairs have a lower excitation energy
than bulk electron-hole pairs. At finite temperature or in a disordered
system where domain walls are created, confined electron-hole pairs are the
lowest-energy excitations but only in a small range of bias.

When confined electron-hole pairs are excited, it should be possible,
according to the theory developed in Ref. \onlinecite{jungwirth} to observe
a breakdown of the quantum Hall effect. In previously studied Ising systems
in semiconductor quantum wells, this breakdown lead to the apparition of
magnetoresistance spikes near the Ising transitions. These spikes were
observed experimentally\cite{poortere}. According to Ref. %
\onlinecite{jungwirth}, when the chemical potential is pinned in the bulk
quasiparticle energy, the confined electron-hole pairs can be easily
excited. If the domain wall loops are dense enough to overlap, these
confined quasiparticles can cross the sample by scattering from one loop to
an adjacent one and then increase the dissipation in the sample, creating
magnetoresistance spikes that are detected near integer fillings close to
the transition between the ordered and the paramagnetic phases at finite
temperature. As far as we know, there has been to date no experimental study
of such effect in the higher Landau levels of bilayer graphene.

The domain wall solution discussed in this work comes from the Coulomb
interaction. It is different from the domain walls studied before in bilayer
graphene\cite{fertig1} which were artificially created in the middle of a
double-gated bilayer graphene sample by placing it in an electric gate where
the potential profile changes sign across the center\cite{martin}. Domain
walls have also recently been studied in bilayer graphene but at zero
magnetic field\cite{li}.

Another type of excitation should be considered in addition to those studied
in this work. It consists of confined textured excitations in the coherent
regions of the domain walls i.e. confined solitons and antisolitons of the
valley pseudospin in the one-dimensional coherent region of a wall. It is
possible to extract an effective pseudospin Hamiltonian for such excitations
by combining the long-wavelength energy functional presented in this paper
with the microscopic Hartree-Fock calculation. This has been done previously
in Ising QHFs\cite{brey,jungwirth, muraki, falko} and in the coherent stripe
phase in a semiconductor double quantum-well system\cite{doiron}. We leave
this problem to future work.

\begin{acknowledgments}
R. C\^{o}t\'{e} was supported by a grant from the Natural Sciences and
Engineering Research Council of Canada (NSERC). Computer time was provided
by Calcul Qu\'{e}bec and Compute Canada.
\end{acknowledgments}

\appendix{}{}

\section{PROJECTED REPRESENTATION IN THE SYMMETRIC GAUGE}

The total density in Landau level $N$ is given by%
\begin{equation}
\widetilde{n}\left( \mathbf{r}\right) =\sum_{\mu }\int dz\left\langle \Psi
_{\mu }^{\dag }\left( \mathbf{r},z\right) \Psi _{\mu }\left( \mathbf{r}%
,z\right) \right\rangle .
\end{equation}%
In the absence of valley coherence, the spin density field in real space is
given by%
\begin{equation}
\widetilde{S}_{z}\left( \mathbf{r}\right) =\frac{1}{2}\sum_{\mu }\mu
_{\sigma }\int dz\left\langle \Psi _{\mu }^{\dag }\left( \mathbf{r},z\right)
\Psi _{\mu }\left( \mathbf{r},z\right) \right\rangle ,
\end{equation}%
and%
\begin{eqnarray}
\widetilde{S}_{x}\left( \mathbf{r}\right) +i\widetilde{S}_{y}\left( \mathbf{r%
}\right) &=&\sum_{\xi }\int dz \\
&&\times \left\langle \Psi _{\xi ,+}^{\dag }\left( \mathbf{r},z\right) \Psi
_{\xi ,-}\left( \mathbf{r},z\right) \right\rangle .  \notag
\end{eqnarray}%
Similarly, in the absence of spin coherence, the valley-pseudospin field
given in real space by%
\begin{equation}
\widetilde{P}_{z}\left( \mathbf{r}\right) =\frac{1}{2}\sum_{\mu }\mu _{\xi
}\int dz\left\langle \Psi _{\mu }^{\dag }\left( \mathbf{r},z\right) \Psi
_{\mu }\left( \mathbf{r},z\right) \right\rangle ,
\end{equation}%
and%
\begin{eqnarray}
\widetilde{P}_{x}\left( \mathbf{r}\right) +i\widetilde{P}_{y}\left( \mathbf{r%
}\right) &=&\sum_{\sigma }\int dz \\
&&\times \left\langle \Psi _{+,\sigma }^{\dag }\left( \mathbf{r},z\right)
\Psi _{-,\sigma }\left( \mathbf{r},z\right) \right\rangle .  \notag
\end{eqnarray}

These density and fields contain the character of the different Landau level
orbitals $n$ in $\Phi _{n,m}\left( \mathbf{r}\right) $ that enter the
definition of the spinors in Eq. (\ref{spinor1},\ref{spinor2}) with the wave
functions in the symmetric gauge. Since these spinors contain a mixture of
the orbitals $n,n-1$ and $n-2,$ the resulting field pattern of a skyrmion in
real space is complex. To get a simpler visualization of the spin and
pseudospin fields, it is useful to use a projected representation that we
define in the following way. We write%
\begin{eqnarray}
\Lambda _{\mu ,\mu ^{\prime }}\left( \mathbf{r}\right) &=&\int dz\Psi _{\mu
}^{\dag }\left( \mathbf{r},z\right) \Psi _{\mu ^{\prime }}\left( \mathbf{r}%
,z\right) \\
&=&\sum_{m_{1},m_{2}}\Gamma _{m_{1},m_{2}}^{\mu _{\xi },\mu _{\xi }^{\prime
}}\left( \mathbf{r}\right) c_{\mu ,m_{1}}^{\dag }c_{\mu ^{\prime },m_{2}}, 
\notag
\end{eqnarray}%
where, for example,%
\begin{eqnarray}
\Gamma _{m_{1},m_{2}}^{+,+}\left( \mathbf{r}\right) &=&b_{+,1}^{\ast
}b_{-,1}\Phi _{n-1,m_{1}}^{\ast }\left( \mathbf{r}\right) \Phi
_{n-1,m_{2}}\left( \mathbf{r}\right) \\
&&+b_{+,2}^{\ast }b_{-,2}\Phi _{n-2,m_{1}}^{\ast }\left( \mathbf{r}\right)
\Phi _{n,m_{2}}\left( \mathbf{r}\right)  \notag \\
&&+b_{+,3}^{\ast }b_{-,3}\Phi _{n,m_{1}}^{\ast }\left( \mathbf{r}\right)
\Phi _{n-2,m_{2}}\left( \mathbf{r}\right)  \notag \\
&&+b_{+,4}^{\ast }b_{-,4}\Phi _{n-1,m_{1}}^{\ast }\left( \mathbf{r}\right)
\Phi _{n-1,m_{2}}\left( \mathbf{r}\right)  \notag
\end{eqnarray}%
and the other components are easily obtained from the spinors in Eqs. (\ref%
{spinor1},\ref{spinor2}). We define the Fourier transform of~$\Gamma
_{m_{1},m_{2}}^{\xi ,\xi ^{\prime }}\left( \mathbf{r}\right) $ as 
\begin{equation}
\Gamma _{m_{1},m_{2}}^{\xi ,\xi ^{\prime }}\left( \mathbf{q}\right) =\int d%
\mathbf{r}\Gamma _{m_{1},m_{2}}^{\xi ,\xi ^{\prime }}\left( \mathbf{r}%
\right) e^{-i\mathbf{q}\cdot \mathbf{r}}.
\end{equation}%
In the symmetric gauge, the matrix elements that enter the Fourier transform
are given by%
\begin{eqnarray}
&&\int d\mathbf{r}\Phi _{n_{1},m_{1}}\left( \mathbf{r}\right) e^{-i\mathbf{q}%
\cdot \mathbf{r}}\Phi _{n_{2},m_{2}}\left( \mathbf{r}\right) \\
&=&\left( \pm i\right) ^{\left\vert n_{1}-n_{2}\right\vert }\left[
F_{n_{1},n_{2}}\left( -\mathbf{q}\right) \right] ^{\ast
}F_{m_{1},m_{2}}\left( -\mathbf{q}\right) ,  \notag
\end{eqnarray}%
with 
\begin{eqnarray}
F_{n_{1},n_{2}}\left( \mathbf{q}\right) &=&\sqrt{\frac{Min\left(
n_{1},n_{2}\right) !}{Max\left( n_{1},n_{2}\right) !}}e^{-\frac{1}{4}%
q^{2}\ell ^{2}} \\
&&\times \left( \frac{\left( \pm q_{y}+iq_{x}\right) \ell }{\sqrt{2}}\right)
^{\left\vert n_{1}-n_{2}\right\vert }L_{Min\left( n_{1},n_{2}\right)
}^{\left\vert n_{1}-n_{2}\right\vert }\left( \frac{q^{2}\ell ^{2}}{2}\right)
,  \notag
\end{eqnarray}%
where the upper sign is for $n_{1}\geq n_{2}$ and the lower sign for $%
n_{1}<n_{2}$ and $L_{n}^{a}\left( x\right) $ is a generalized Laguerre
polynomial$.$ It follows that 
\begin{equation}
\Gamma _{m_{1},m_{2}}^{+,+}\left( \mathbf{q}\right) =\Omega _{+,+}\left( -%
\mathbf{q}\right) F_{m_{1},m_{2}}\left( \mathbf{q}\right) ,
\end{equation}%
where $\Omega _{+,+}\left( \mathbf{q}\right) $ does not involve the quantum
number $m$ and is given by 
\begin{eqnarray}
\Omega _{+,+}\left( \mathbf{q}\right) &=&b_{+,1}^{\ast }b_{-,1}\left[
F_{n-1,n-1}\left( \mathbf{q}\right) \right] ^{\ast } \\
&&-b_{+,2}^{\ast }b_{-,2}\left[ F_{n-2,n}\left( \mathbf{q}\right) \right]
^{\ast }  \notag \\
&&-b_{+,3}^{\ast }b_{-,3}\left[ F_{n,n-2}\left( \mathbf{q}\right) \right]
^{\ast }  \notag \\
&&+b_{+,4}^{\ast }b_{-,4}\left[ F_{n-1,n-1}\left( \mathbf{q}\right) \right]
^{\ast }.  \notag
\end{eqnarray}%
We thus get%
\begin{eqnarray}
\Lambda _{\mu ,\mu ^{\prime }}\left( \mathbf{q}\right)
&=&\sum_{m_{1},m_{2}}\Omega _{\mu _{\xi },\mu _{\xi }^{\prime }}\left( -%
\mathbf{q}\right) F_{m_{1},m_{2}}\left( -\mathbf{q}\right) \\
&&\times c_{\mu ,m_{1}}^{\dag }c_{\mu ^{\prime },m_{2}}.  \notag
\end{eqnarray}


\begin{thebibliography}{99}
\bibitem{jung} The value of the hopping parameters that we use are those
recently computed by Jeil Jung and Allan H. MacDonald, Phys. Rev. B \textbf{%
89}, 035405 (2014). Our definition of the parameters $\gamma _{0},\gamma
_{1},\gamma _{4}$ differs by a minus sign from that used by these authors.
\bibitem{note4} This is true in the gradient approximation but not in the
microscopic Hartree-Fock approach unless the state is uniform or modulated
in one direction of space only.
\bibitem{note3} In Ref. \onlinecite{luo}, this is incorrectly written as $%
C_{n,m}=1$ if $m<n$ and $C_{n,m}=\left( -1\right) ^{m-n}$ if $m\geq n.$
\end{thebibliography}
\end{document}